\documentclass[prl,reprint,amsmath,amssymb,amsfonts,superscriptaddress]{revtex4-2}
\usepackage[T1]{fontenc}
\usepackage{mathptmx}
\usepackage{bm}
\usepackage{color,graphicx}
\usepackage[version=4]{mhchem}
\usepackage{mathrsfs}
\usepackage{siunitx}
\usepackage{caption}
\usepackage{pdfpages}
\usepackage{xcolor,hyperref}
\hypersetup{
   colorlinks,
   linkcolor={red!80!black},
   citecolor={blue!50!black},
   urlcolor={blue!80!black}
}

\makeatletter
\AtBeginDocument{\let\LS@rot\@undefined}
\makeatother

\newcommand{\diff}{\mathrm{d}}

\begin{document}
%\setpagewiselinenumbers
%\modulolinenumbers[5]
%\linenumbers

	\title{Ion steric effect induces giant enhancement of thermoelectric conversion in electrolyte-filled nanochannels}
\author{Wenyao Zhang}
\affiliation{MOE Key Laboratory of Thermo-Fluid Science and Engineering, School of Energy and Power Engineering, Xi'an Jiaotong University, Xi'an 710049, China}
\author{Xinxi Liu}
\author{Kai Jiao}
\affiliation{MOE Key Laboratory of Thermo-Fluid Science and Engineering, School of Energy and Power Engineering, Xi'an Jiaotong University, Xi'an 710049, China}
\author{Qiuwang Wang}
\affiliation{MOE Key Laboratory of Thermo-Fluid Science and Engineering, School of Energy and Power Engineering, Xi'an Jiaotong University, Xi'an 710049, China}
\author{Chun Yang}
%\email{mcyang@ntu.edu.sg}
\affiliation{School of Mechanical and Aerospace Engineering, Nanyang Technological University, 50 Nanyang Avenue, Singapore 639798, Singapore}
\author{Cunlu Zhao}
\email{mclzhao@xjtu.edu.cn}
\affiliation{MOE Key Laboratory of Thermo-Fluid Science and Engineering, School of Energy and Power Engineering, Xi'an Jiaotong University, Xi'an 710049, China}

\begin{abstract}
	Ionic thermoelectricity in nanochannels has received increasing attention because of its advantages such as high Seebeck coefficient and low cost. However, most studies have focused on dilute simple electrolytes that neglect the effects of finite ion sizes and short-range electrostatic correlation. Here, we reveal a new thermoelectric mechanism arising from the coupling of ion steric effect due to finite ion sizes and ion thermodiffusion in electric double layers, using both theoretical and numerical methods. We show that this mechanism can significantly enhance the thermoelectric response in nanoconfined electrolytes, depending on the properties of electrolytes and nanochannels. Compared to the previously known mechanisms, the new mechanism can increase the Seebeck coefficient by 100\% or even one order of magnitude enhancement under optimal conditions. Moreover, we demonstrate that the short-range electrostatic correlation can help preserve the Seebeck coefficient enhancement in weaker confinement or in more concentrated electrolytes.
\end{abstract}

\maketitle

Seebeck effect refers to a phenomenon in which an electromotive force develops in electrically conducting materials subject to an externally applied temperature difference \cite{Bell2008,Shi2020}. Diverse technologies mainly take advantage of the Seebeck effect in two ways, namely, temperature sensing \cite{Herwaarden1986} and direct (solid-state) thermoelectric conversion \cite{Ichinose2019,HurtadoGallego2022}. Nowadays, the direct thermoelectric energy conversion has become an emerging technology to harvest the low-grade thermal energy ($<\SI{100}{\degreeCelsius}$) that is ubiquitous and abundant in nature and industrial processes \cite{Forman2016,Venkatasubramanian2019}, and thus holds the potential for reducing carbon emission. However, traditional solid-state thermoelectric materials/devices (such as \ce{Bi2Te3}) are not only expensive but also have become low-efficient to harvest the low-grade thermal energy \cite{Yu2020}. In comparison, the ionic Seebeck effect provides a promising choice since it has advantages of high Seebeck coefficient and low cost \cite{dietzel2016,Li2019,Fu2019}.

In the presence of a temperature gradient $\nabla T$, separation of charge carriers (i.e., dissolved ion species) in liquid electrolytes can give rise to a thermoelectric field \cite{Wurger_2010} ${\bm E}_{\infty}={k_{\mathrm{B}}(\alpha_+-\alpha_-)}\nabla T/(ez)$ due to the difference in the reduced Soret coefficients between cation ($\alpha_+$) and anion ($\alpha_-$); here $k_{\rm B}$ is the Boltzmann constant, $e$ the elementary charge and $z$ the ion valence. This thermoelectric mechanism occurs in bulk electrolytes without confinement, and is fundamentally the same as that of semiconductors-based thermoelectricity except in charge carriers. In this case, the electrolyte concentration $n$ in the bulk solution reaches the classic Soret equilibrium which can be mathematically described by \cite{Wurger_2010}:
\begin{equation}\label{eq:soret_eq}
	\frac{\nabla n}{n} = - \frac{\alpha_++\alpha_-}{T}\nabla T\equiv - \alpha \frac{\nabla T}{T}
\end{equation}

Unfortunately, the ionic thermoelectric energy conversion in liquid electrolytes has not received enough attention until high Seebeck coefficients were experimentally observed in a number of unconfined electrolytes \cite{Bonetti2011,Bonetti2015,Zhao2016} and the Seebeck coefficient in liquid electrolytes was shown to be enhanced by nanoconfinement \cite{dietzel2016}. A series of subsequent investigations made further detailed characterization of thermoelectric transport behavior in solutions of simple electrolytes in nanopores, and the enhancement of the Seebeck coefficient due to nanoconfinement was widely confirmed \cite{Dietzel2017,Zhang2019,Fu2019,Zhong2020,Zhong2020a,zhang2022sim,Qian2022confinement}. In these studies, the nanochannel wall is electrically charged and accordingly an electric double layer (EDL) develops near the channel wall \cite{Schoch2008,Bocquet2010}. When subject to a temperature gradient, the local EDL structure is modified to induce an extra thermoelectric mechanism which contributes to the enhancement of Soret coefficient \cite{dietzel2016}. However, the studies mentioned above focused on ionic thermoelectricity of simple liquid electrolytes (such as NaCl and KCl under dilute or low ionic strength conditions), and overlooked the steric effect due to the finite ion sizes \cite{Borukhov1997,Tessier2006} and the short-range electrostatic correlation effect as well as their contributions to the thermoelectric performance \cite{Bazant2009,Bazant2011}. These overlooked effects, however, are particularly prominent in complex electrolytes which have high ionic strength or large ion size, or contact highly charged surfaces \cite{Li2019,Huang2015}.

Here, we report a new thermoelectric mechanism in confined electrolytes caused by the joint action of the steric effect due to the finite ion sizes and ion thermodiffusion in the EDL using both theoretical analysis and numerical simulation. We further show that this new mechanism can significantly enhance the thermoelectric response by (at least) several times, and the short-range interionic electrostatic correlation can help to maintain the enhancement in a larger range of confinement size or electrolyte concentration.

%%%%%%%%%%%%%%%%  FIGURE 1 %%%%%%%%%%%%%%%%%%%%%
\begin{figure*}[t]
\centering
\includegraphics{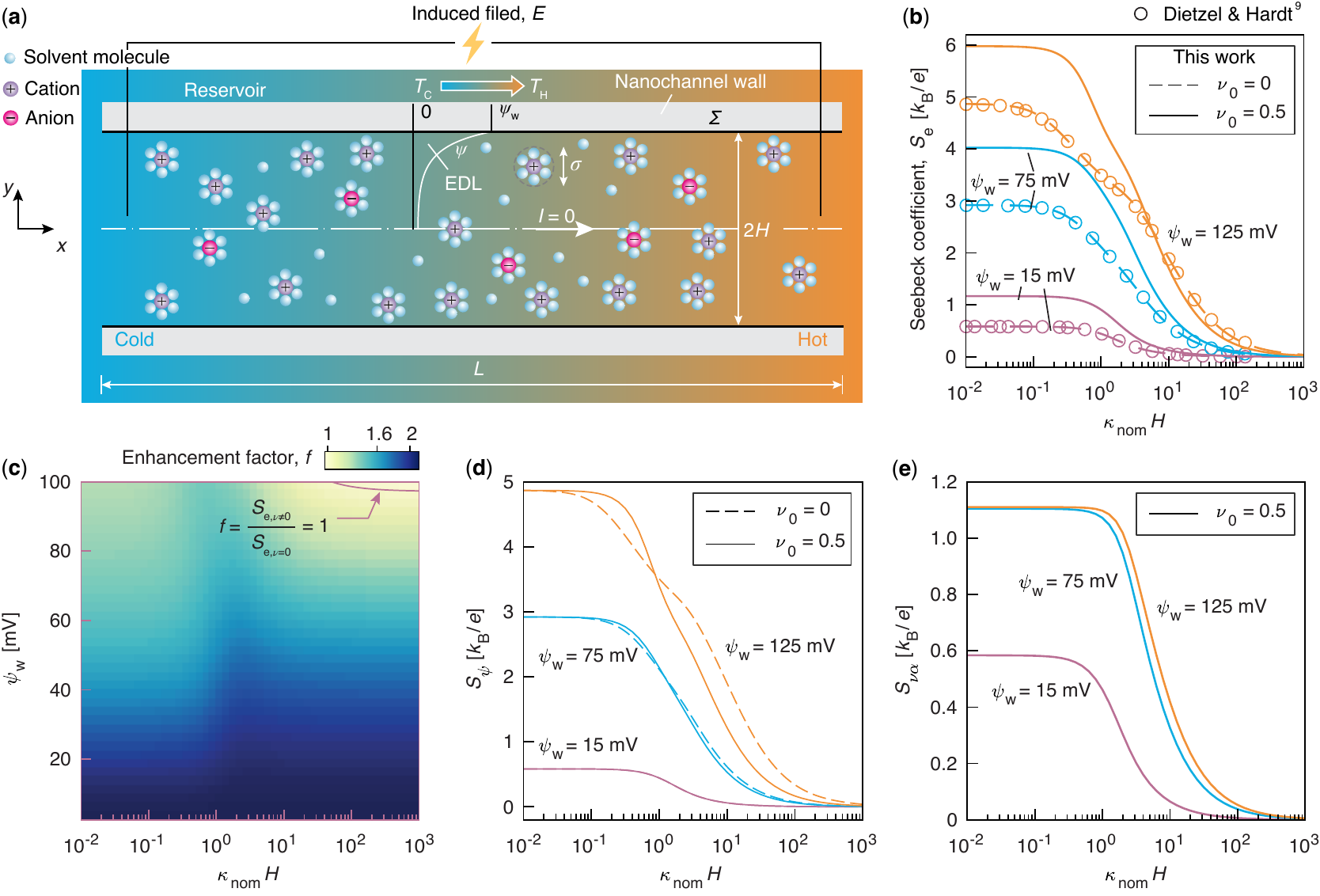}
\caption{{Ion steric effect on thermoelectric conversion}. (a) Schematic of a slit nanochannel of width $2H$ and length $L$, submerged in a big reservoir containing an electrolyte with a bulk number concentration of $n_0$ and subjected to a temperature difference of $\Delta T= T_{\mathrm{H}}-T_{\mathrm{C}}$. The nanochannel wall is kept at a constant surface potential of $\psi_{\rm w}$ or bears a constant charge density of $\varSigma$. (b) Seebeck coefficient as a function of nominal Debye parameter $\kappa_{\mathrm{nom}} H$ for varying $\psi_{\mathrm{w}}$ values as $\nu_0 =0$ (dash lines) and $\nu_0=0.1$ (solid lines). Symbols are taken from ref.~\cite{dietzel2016} and lines are computed by Eq.~(\ref{eqn:Se}). (c) Enhancement factor $S_{{\rm e},\nu\neq 0}/S_{{\rm e},\nu=0}$ as a function of $\kappa_{\rm nom} H$ and $\psi_{\rm w}$. (d) Seebeck coefficient associated with the temperature dependency of ion electrophoretic mobilities, $S_{\psi}$, as a function of $\kappa_{\rm nom}H$ for varying $\psi_{\rm w}$ as $\nu_0 =0$ (dash lines) and $\nu_0=0.1$ (solid lines). (e) Seebeck coefficient associated with the ion steric effect, $S_{\nu\alpha}$, as a function of $\kappa_{\rm nom}H$ for varying $\psi_{\rm w}$ as $\nu_0=0.1$. The reduced Soret coefficients of ions were set to $\alpha_{+}=\alpha_{-}=5$ and the normalized difference in ion diffusion coefficients was set to $\chi=0$. All results were calculated at $T_{0}=(T_{\rm C}+T_{\rm H})/2=298$ K.}
\label{fig:fig1}
\end{figure*}

We consider a system comprising of an electrolyte-filled nanochannel with a width of $2H$ and a length of $L$ (Fig.~\ref{fig:fig1}a). The wall temperature of the nanochannel linearly increases from $T_{\rm C}$ to $T_{\rm H}$ with a temperature difference of $\Delta T = T_{\rm H} - T_{\rm C}$. In addition, all dissolved ions are assumed to have the same effective diameters $\sigma$. The nanochannel provides a promising platform for energy conversion \cite{Heyden2006,Siria2013,Feng2016,Zhang2021} and fluid pumping \cite{Ritt2022}. Note that the longitudinal (i.e., $x$-direction) confinement effect would weaken the Seebeck coefficient \cite{Qian2022confinement,wurger2020thermopower}. Thus, in this study we only consider the case without the longitudinal confinement effect but with the lateral (i.e., $y$-direction) confinement effect due to the nanochannel. This implies $\delta^2=(H/L)^2 \ll 1$ and facilitates the upcoming lubrication analysis. With order-of-magnitude analysis~\cite{dietzel2016,Zhang2019}, we only need to consider the conductive heat transfer inside the nanochannel and obtain $\nabla\cdot(k\nabla T)=0$, where $k$ is the thermal conductivity of the solution and $T$ is the absolute temperature. Therefore, to the first order in $\delta$, a linear temperature profile, $T=T_{\rm C}+(x/L)\Delta T$, is shown to build up along the axial direction of the nanochannel \cite{dietzel2016} and is also confirmed by numerical simulation.

For the system shown in Fig.~\ref{fig:fig1}a, the ionic flux $\bm{J}_{i}$ is governed by the modified Nernst-Planck equation (Sec. S1, Supporting Information)
\begin{equation}\label{eq:ionflux}
	\bm{J}_{i} = -D_i \left( \nabla n_i  + \frac{\sigma^3 n_i  \sum_j \nabla n_j}{1-\sigma^3\sum_j n_j} + \frac{z_i e n_i}{k_{\mathrm{B}} T} \nabla \phi + \frac{2 n_i \alpha_i}{T}\nabla T \right)
\end{equation}
where $D_i$, $\alpha_i$, $z_i$ are the diffusion coefficient, the reduced Soret coefficient \cite{Qian2022confinement} and the valence of ion species $i$, respectively. In addition, $\phi$ is the overall electric potential.

The overall electric field $\nabla\phi=\nabla{\psi}-{\bm E}$ can be divided into two parts \cite{Fair1971,Peters2016}: (1) EDL field $\nabla\psi$; (2) Induced electric field ${\bm E}$. Assuming the EDL potential $\psi$ satisfies the local Poisson equation (which will be confirmed by numerical simulation in the following) and thus one can readily find ${\bm E} \approx (E,0)$, which is in line with the analysis given in Ref.~\cite{dietzel2016}.

Following the approach detailed in Refs.~\cite{dietzel2016,Alizadeh2017}, we find that for symmetry $z:z$ electrolytes, the ion concentration satisfies the ``Fermi-like'' distribution  (Sec. S2, Supporting Information):
\begin{equation}\label{eq:mBolt}
	n_{i}(x,y)=\frac{n_{\rm v} \exp\left(- \frac{z_i e \psi}{k_{\mathrm{B}}T} \right)}{1 - 2\sigma^3 n_{\rm v} + 2\sigma^3 n_{\rm v} \cosh \left( \frac{z e \psi}{k_{\mathrm{B}}T} \right)}
\end{equation}
where $n_{\rm v}$ is the concentration of either ionic species in a virtual electroneutral reservoir ($\psi=0$) that is in equilibrium with any cross section of the nanochannel \cite{Baldessari2008,Peters2016,Alizadeh2017} and the local virtual concentration, $n_{\rm v}$, is a function of $x$ and can be determined from Eq.~(\ref{eq:soret_eq}) with the constraint of $\int_0^L n_{\rm v} \diff x = n_0 L $.

Substituting Eq.~(\ref{eq:mBolt}) into Eq.~(\ref{eq:ionflux}), we can obtain the axial ion fluxes as
\begin{align}\label{eq:jx}
	-\frac{J_{\pm,x}}{n_{\pm}D_{\pm}} &= \frac{\diff_x\ln n_{\rm v}}{1-2\sigma^3 n_{\rm v}} + \left( \frac{2\alpha_{\pm}}{T} \pm \frac{ez\psi}{k_{\mathrm{B}}T^2}  \right) \diff_x T\mp \frac{ezE}{k_{\mathrm{B}}T}
\end{align}
Equations (\ref{eq:mBolt}) and (\ref{eq:jx}) have included the steric effect in the double layer beyond the Stern layer.

In the absence of external electric load, the induced thermoelectric field, $E$, is determined by setting the overall electric current, $I=e\int_{-H}^{H} \diff y  (z_{+}J_{+,x}+z_{-}J_{-,x}) $, to zero \cite{Wurger_2010,dietzel2016,Zhang2019,zhang2022sim}. Accordingly, the Seebeck coefficient is expressed as $S_{e} = E/\diff_x T $, and can be evaluated as (Sec. S3, Supporting Information)
\begin{equation}\label{eqn:Se}
	S_e=S_{\delta\alpha} + S_{\psi} +  S_{\nu\alpha}
\end{equation}
with
\begin{equation}\label{eq:Se_delta_alpha}
	S_{\delta\alpha}= \frac{k_{\mathrm{B}}\delta\alpha}{ez}
\end{equation}
\begin{equation}\label{eq:Se_psi}
	S_{\psi} = \frac{k_{\mathrm{B}}}{ez} \frac{\int_{0}^{H} \tilde{\psi}\frac{\cosh(\tilde{\psi})-\chi\sinh(\tilde{\psi})}{1-\nu+\nu \cosh(\tilde{\psi})} \diff y}{\int_{0}^{H} \frac{\cosh(\tilde{\psi})-\chi\sinh (\tilde{\psi})}{1-\nu+\nu \cosh(\tilde{\psi})} \diff y}
\end{equation}
\begin{equation}\label{eq:Se_nu}
	S_{\nu\alpha} =  \frac{k_{\mathrm{B}}}{ez} \frac{\nu \alpha}{1-\nu} \frac{\int_{0}^{H} \frac{\sinh(\tilde{\psi}) - \chi \cosh (\tilde{\psi})}{1-\nu+\nu \cosh(\tilde{\psi})} \diff y}{\int_{0}^{H} \frac{\cosh(\tilde{\psi}) - \chi\sinh (\tilde{\psi})}{1-\nu+\nu \cosh(\tilde{\psi})} \diff y}
\end{equation}
where $\tilde{\psi}=ez\psi/(k_{\mathrm{B}}T)$, $\chi=(D_{+}-D_{-})/(D_{+}+D_{-})$ and $\delta\alpha=\alpha_{+}-\alpha_{-}$. In addition, $\nu = 2\sigma^3 n_{\rm v} $ is the volume fraction of solvated ions in the virtual reservoir, which characterizes the degree of ion crowding and the importance of steric effect \cite{Bazant2011}. The higher $\nu$ is, the more important the steric effect becomes. For typical ions with hydration shells like \ce{Na+} and \ce{Cl-}, $\sigma\sim \SI{7}{\angstrom}$~\cite{Volkov1997}, thus $\nu$ can reach $\sim 0.4$ at a salt concentration of \SI{1}{M} (Fig. S1). Higher $\nu$ values (around and above 0.5) are expected for ionic liquids with larger ion diameters or higher concentrations~\cite{Bazant2011,Fedorov2014,Gebbie2015}. Hence, it is safe to consider the cases in which $\nu$ is as high as $\sim 0.1$.

Equation~(\ref{eqn:Se}) constitutes the key result of the present work and clearly indicates three thermoelectric mechanisms of nanoconfined electrolytes: $S_{\delta\alpha}$ is well-known and originates from the difference in the Soret coefficients between cations and anions \cite{Wurger_2010}; $S_{\psi}$ was discovered in 2016 and originates from the temperature dependencies of the ion electrophoretic mobilities \cite{dietzel2016}; $S_{\nu\alpha}$ is the new thermoelectric mechanism reported in this study and originates from the coupling of the ion steric effect and ion thermodiffusion in the EDL.

To evaluate $S_{e}$, we need to determine $\psi$ first. Substituting Eq.~(\ref{eq:mBolt}) into the Poisson equation $\nabla\cdot ( \epsilon\nabla \psi) = -ez(n_{+}-n_{-}) $ (here $\varepsilon=\varepsilon(T(x))$ is the dielectric permittivity of the liquid electrolytes and $\partial_y\varepsilon=0$) and neglecting all terms of order of $\delta^2$ (including $\partial_{xx}\psi$ and $\partial_x\varepsilon\partial_x\psi$ terms), we obtain
\begin{equation}\label{eq:pfe}
	\frac{\partial^2 \tilde{\psi}}{\partial y^2} \approx \frac{\kappa^2 \sinh (\tilde{\psi}) }{1-\nu +\nu \cosh ( \tilde{\psi} )}
\end{equation}
which is the so-called Poisson-Fermi equation. Herein, $\kappa = \sqrt{2e^2z^2n_{\rm v}/(\epsilon k_{\rm B} T)}$ is the reciprocal of the local Debye length characterizing the EDL thickness. Since both $\kappa$ and $\nu$ are functions of $x$, we can deduce that $\tilde{\psi}=\tilde{\psi}(x,y)$. For convenience, we set $\kappa_{\rm nom} = \kappa(x=L/2)$ and $\nu_0=\nu(x=L/2)$ as references with $ n_{\rm v}=n_0$ and $ T=T_0$. Then, Eq.~(\ref{eq:pfe}) can be solved numerically with either the fixed surface potential boundary condition, i.e., $\tilde{\psi}(y=H)=ez\psi_{\rm w}/(k_{\rm B}T)$ or the fixed surface charge density boundary condition, i.e., $\varepsilon\partial_y \tilde{\psi}(y=H)=e\varSigma/(k_{\rm B}T)$.

We first compare the Seebeck coefficients with and without the ion steric effect for varying $\kappa_{\rm nom}H$ (which characterizes the ratio of $H$ to nominal Debye length $\kappa_{\rm nom}^{-1}$) and $\psi_{\rm w}$. Here, we set $\alpha_+=\alpha_-=5$ (Sec. S6, Supporting Information) to discard the contribution of the conventional bulk thermoelectric mechanism, namely, $S_{\delta\alpha}$ given by Eq.~(\ref{eq:Se_delta_alpha}). It is observed that the inclusion of the ion steric effect can enhance the Seebeck coefficient except for the cases with high surface potential and large $\kappa_{\rm nom}H$ (i.e., weak confinement, Fig.~\ref{fig:fig1}b). To evaluate such enhancement effect, we introduce an enhancement factor defined as $f=S_{{\rm e},\nu\neq 0}/S_{{\rm e},\nu=0}$ (Fig.~\ref{fig:fig1}c). For low $\psi_{\rm w}$ values, the enhancement factor $f\sim 2$, which indicates that the Seebeck coefficients with ion steric effect can double compared to those without ion steric effect. As $\psi_{\rm w}$ increases, one finds that $f$ decreases regardless of $\kappa_{\rm nom}H$ values, implying that the enhancement effect gradually diminishes. For high $\psi_{\rm w}$ values, $f$ even decays to below 1 as long as $\kappa_{\rm nom}H$ increases above a critical value. This implies that the Seebeck coefficient is weakened by the ion steric effect. This is because the enhancement in $S_{\rm e}$ due to $S_{\nu\alpha}$ is counteracted by $S_{\psi}$ induced the reduction of the Seebeck coefficient (Fig.~\ref{fig:fig1}d), which arises from the decrease of surface charge density due to the ion steric effect \cite{Hatlo2012} (Table S1).

%%%%%%%%%%%%%%%%  FIGURE 2 %%%%%%%%%%%%%%%%%%%%%
\begin{figure*}[t]
\centering
\includegraphics{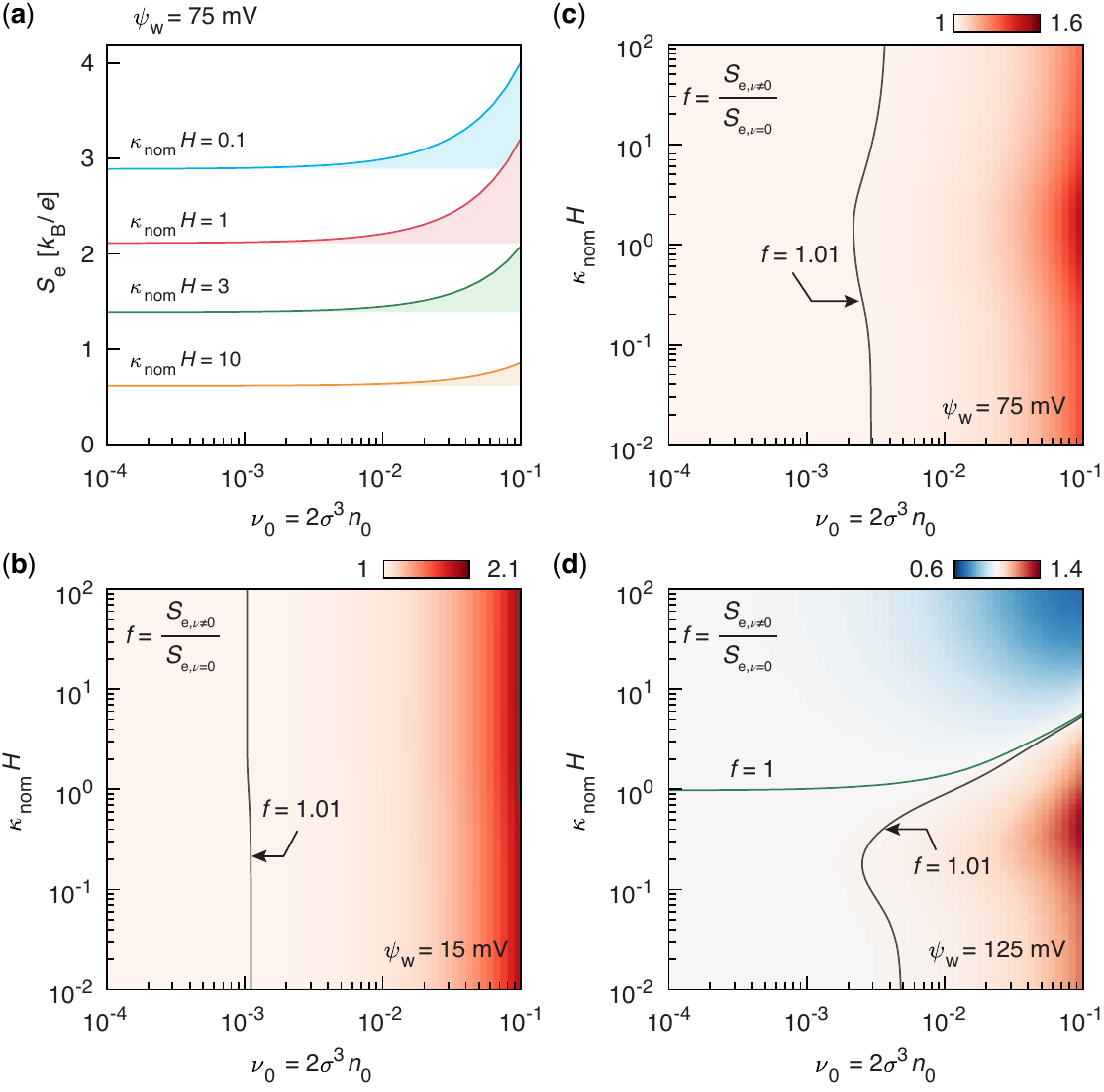}
\caption{Effect of ion volume fraction on thermoelectric conversion. (a) Seebeck coefficient as a function of ion volume fraction $\nu_0$ for varying $\kappa_{\mathrm{nom}} H$ values as $\psi_{\mathrm{w}} = 75\ $mV$ $. The filled areas stand for the ion-steric-effect ($\nu_0 \neq 0$) induced enhancement in the Seebeck coefficient compared with the case of $\nu_0 =0$. (b-d) Enhancement factor, $f=S_{{\rm e},\nu\neq 0}/S_{{\rm e},\nu=0}$ as a function of $\kappa_{\rm nom}H$ and $\nu_0$ for $\psi_{\mathrm{w}} = \SI{15}{mV}  $ (b), \SI{75}{mV} (c) and \SI{125}{mV} (d). Other parameter values are the same as those in Fig.~\ref{fig:fig2}.}
\label{fig:fig2}
\end{figure*}

Next, we need to answer such a question: how does the parameter measuring the ion steric effect, $\nu_0$, affect the Seebeck coefficient? Clearly, for $\psi_{\rm w}=\SI{75}{mV}$ as $\nu_0 \lesssim 3 \times 10^{-3} $, increasing $\nu_0$ almost has no effect on the Seebeck coefficient (Fig.~\ref{fig:fig2}a, c). 
As $\nu_0$ increases beyond $\sim 3\times 10^{-3}$, the enhancement factor, $f=S_{{\rm e},\nu\neq 0}/S_{{\rm e},\nu= 0}$, starts to surpass unity by more than 1\% (Fig.~\ref{fig:fig2}c). 
As $\nu_0$ increases to 0.1, $f$ can reach ca. 1.6 at $\kappa_{\rm nom}H\sim 1$. The similar trend can also be observed for $\psi_{\rm w}=\SI{15}{mV}$ (Fig.~\ref{fig:fig2}b). 
In comparison, in this case $f$ begins to surpass 1 by more than 1\% at smaller $\nu_0\sim 10^{-3}$ and $f$ can increase beyond 2 as $\nu_0 \gtrsim 0.09 $ and $\kappa_{\rm nom}H \gtrsim 0.1 $. 
However, for $\psi_{\rm w}=\SI{125}{mV}$, the behavior is quite different (Fig.~\ref{fig:fig2}d). In this case, for all $\nu_0$ considered, the enhancement in $S_{\rm e}$ (i.e., $f>1$) only occurs at relatively small $\kappa_{\rm nom}H$ (whose range is dependent on $\nu_0$, see the lower right of Fig.~\ref{fig:fig2}d), while if $\kappa_{\rm nom}H$ surpasses certain values (dependent on $\nu_0$), $S_{\rm e}$ is even weakened by the ion steric effect (i.e., $f<1$, also seen in Fig.~\ref{fig:fig2}a). {Moreover, we observe that for cases of $\kappa_{\rm nom}H=O(1)$, $f$ surpasses unity by 1\% at relatively smaller $\nu_0$ compared with cases of smaller or larger $\kappa_{\rm nom}H$}. 

In addition, we should emphasize that the thermoelectric conversion due to the coupling of the ion steric effect and the ion thermodiffusion in EDLs (denoted by $S_{\nu\alpha}$) is a confinement effect when $\chi=0$ (which does hold as $\chi\neq 0$, see Sec. S3), in analogy to that due to the temperature dependency of ion electrophoretic moblities~\cite{dietzel2016} (denoted by $S_{\psi}$). This is directly demonstrated by Fig.~\ref{fig:fig1}e, in which we observe that as $\kappa_{\rm nom}H\rightarrow 0$  (i.e., $H\ll\kappa_{\rm nom}^{-1}$), $S_{\nu\alpha}$ arrives at its maximum value, while as $\kappa_{\rm nom}H\rightarrow\infty$ (i.e., $H\gg\kappa_{\rm nom}^{-1}$), $S_{\nu\alpha}$ decays to zero. Notably, for high $\psi_{\rm w}$ values, $S_{\nu\alpha}$ saturates to a plateau value as $\kappa_{\rm nom}H\rightarrow 0$. For highly charged surfaces (i.e., $|\tilde{\psi}| \gg 1$) with heavily overlapped EDLs (i.e., $\kappa_{\rm nom}H\rightarrow 0$), the uniform potential model is valid~\cite{Biesheuvel2011,Peters2016} and $\sinh (\tilde{\psi}) \approx \pm \cosh (\tilde{\psi}) $, thus Eq.~(\ref{eq:Se_nu}) reduces to
\begin{equation}\label{eq:S_nu0}
	|S_{\nu\alpha}|_{|\tilde{\psi}| \gg 1,\ \kappa_{\rm nom}H \rightarrow 0} = \frac{k_{\mathrm{B}}}{ez} \frac{\nu \alpha}{1-\nu}
\end{equation}
which corresponds to the aforementioned plateau value.

%%%%%%%%%%%%%%%%  FIGURE 3 %%%%%%%%%%%%%%%%%%%%%
\begin{figure}[t]
\centering
\includegraphics{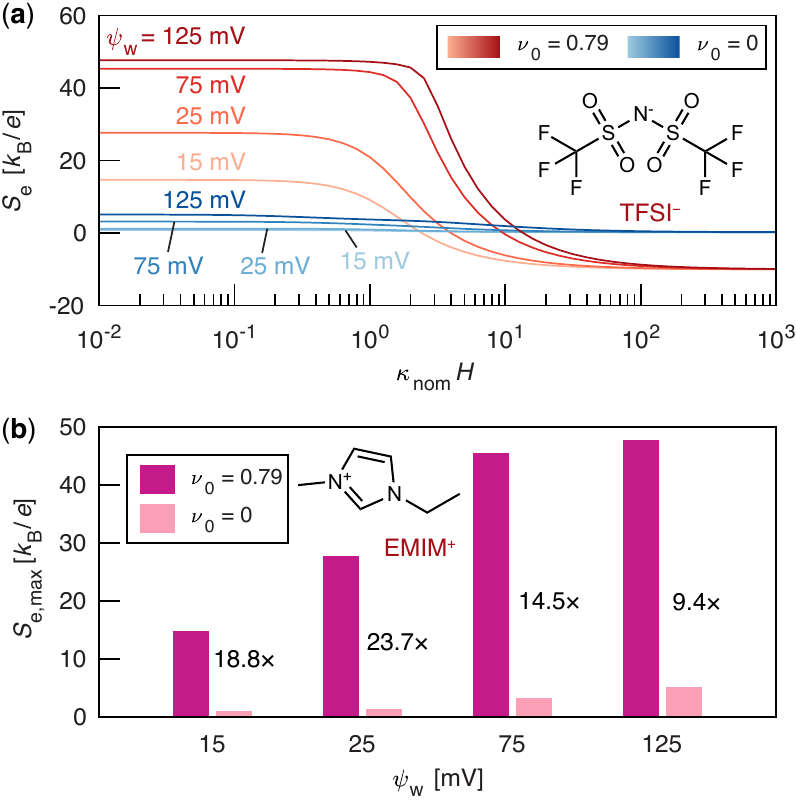}
\caption{Significant enhancement of thermoelectric response in ionic liquids. (a) Overall Seebeck coefficient as a function of $\kappa_{\rm nom}H$ for varying ion volume fractions and varying surface potentials. (b) Maximum Seebeck coefficient as a function of $\kappa_{\rm noom}H$ for varying ion volume fractions and varying surface potentials. In the calculation, the ionic liquid 1-ethyl-3-methylimidazolium bis((trifluoromethyl)sulfonyl)imide [\ce{EMIM^+TFSI-}] is considered. The reduced Soret coefficients of cations and anions are estimated by \cite{Wuerger2021} $\alpha_i=0.5+\Delta H_i/(2k_{\rm B}T)$ with $\Delta H_+=0.27$ eV and $\Delta H_-=0.26$ eV \cite{DAgostino2018}. In addition, $\chi$ is estimated by the ion diffusivity ratio $D_+/D_-=1.64$ \cite{DAgostino2018}. All results were calculated at $T_{0}=(T_{\rm C}+T_{\rm H})/2=298$ K.}
\label{fig:fig3}
\end{figure}

Notably, Eq.~(\ref{eq:Se_nu}) or Eq. (\ref{eq:S_nu0}) suggests that the larger $\alpha$ or $\nu$ is, the higher $S_{\nu\alpha}$ becomes (and so does $S_{\rm e}$, see Fig. \ref{fig:fig3}a). Generally, for electrolytes like ionic liquids $\alpha$ can be related to the ion heat of transport $Q_i$ by $\alpha=\sum_i Q_i/(2k_{\rm B}T)$ \cite{Wurger_2010}, where $Q_i$ is given by $Q_i = k_{\rm B}T + \Delta H_{i}$ \cite{Wuerger2021} with $\Delta H_i$ being the ion activation enthalpy. Taking the ionic liquid \ce{[EMIM]^+[TFSI]-} (see Fig. S2 for its chemical structure and dimensions) as an example, $\alpha = 1+\sum_i\Delta H_i/(2k_{\rm B}T)$ is evaluated as $\sim 11.3$ and $\delta\alpha\sim 0.19$ by noting that $\Delta H_+=\SI{0.27}{eV}$ and $\Delta H_-=\SI{0.26}{eV}$ \cite{DAgostino2018}. Fig.~\ref{fig:fig3}a reveals that the Seebeck coefficient is significantly enhanced for cases with the ion steric effect ($\nu_0=0.79$, see Sec. S3 in Supporting Information) compared to cases without ion steric effect ($\nu_0=0$) at small $\kappa_{\rm nom}H$. The maximum values of $S_{\rm e}$ (being denoted by $S_{\rm e,\max}$), for $\psi_{\rm w}=[15,\ 25,\ 75,\ 125]\ \SI{}{mV}$, are estimated to be $\sim [14.6,\ 27.6,\ 45.3,\ 47.6]k_{\rm B}/e = [1.26,\ 2.38,\ 3.9,\ 4.1]$ \SI{}{mV.K^{-1}}, which correspond to [18.8, 23.7, 14.5, 9.4] times of $S_{\rm e,\max}$ at $\nu_0=0$ (i.e. $\psi_{\rm w}/T_0+\delta\alpha$ \cite{dietzel2016}, see Fig.~\ref{fig:fig3}b). This estimation demonstrates that the new thermoelectric mechanism arising from the coupling of the ion steric effect and ion thermodiffusion in the EDLs can enhance the Seebeck coefficient by several times and even one order of magnitude.

It is worth emphasizing that room-temperature ionic liquids are frequently considered as strongly dissociated and highly concentrated electrolytes and thus typical values of $\nu_0$ can be above 0.5 \cite{Fedorov2014}. For an ionic strength as high as $\sim \SI{1}{M}$, $\kappa_{\rm nom}^{-1}$ is estimated to be $\sim$\SI{1}{\angstrom} with $\varepsilon_{\rm r}=12.3$ (ref.~\cite{Gebbie2015}) and thus $\kappa_{\rm nom}H\geq 10$ if the typical channel height $H\geq \SI{1}{nm}$. It seems that the maximum Seebeck coefficient is unachievable since practical values of $\kappa_{\rm nom}H$ would be much larger than unity \cite{Bazant2011}. However, in ionic liquids the ion correlation becomes practically important and must be taken into accont, and later we would show that the maximum Seebeck coefficient can be achievable at much larger $\kappa_{\rm nom}H$ by considering the electrostatic correlation.

%%%%%%%%%%%%%%%%  FIGURE 4 %%%%%%%%%%%%%%%%%%%%%
\begin{figure*}[t]
\centering
\includegraphics{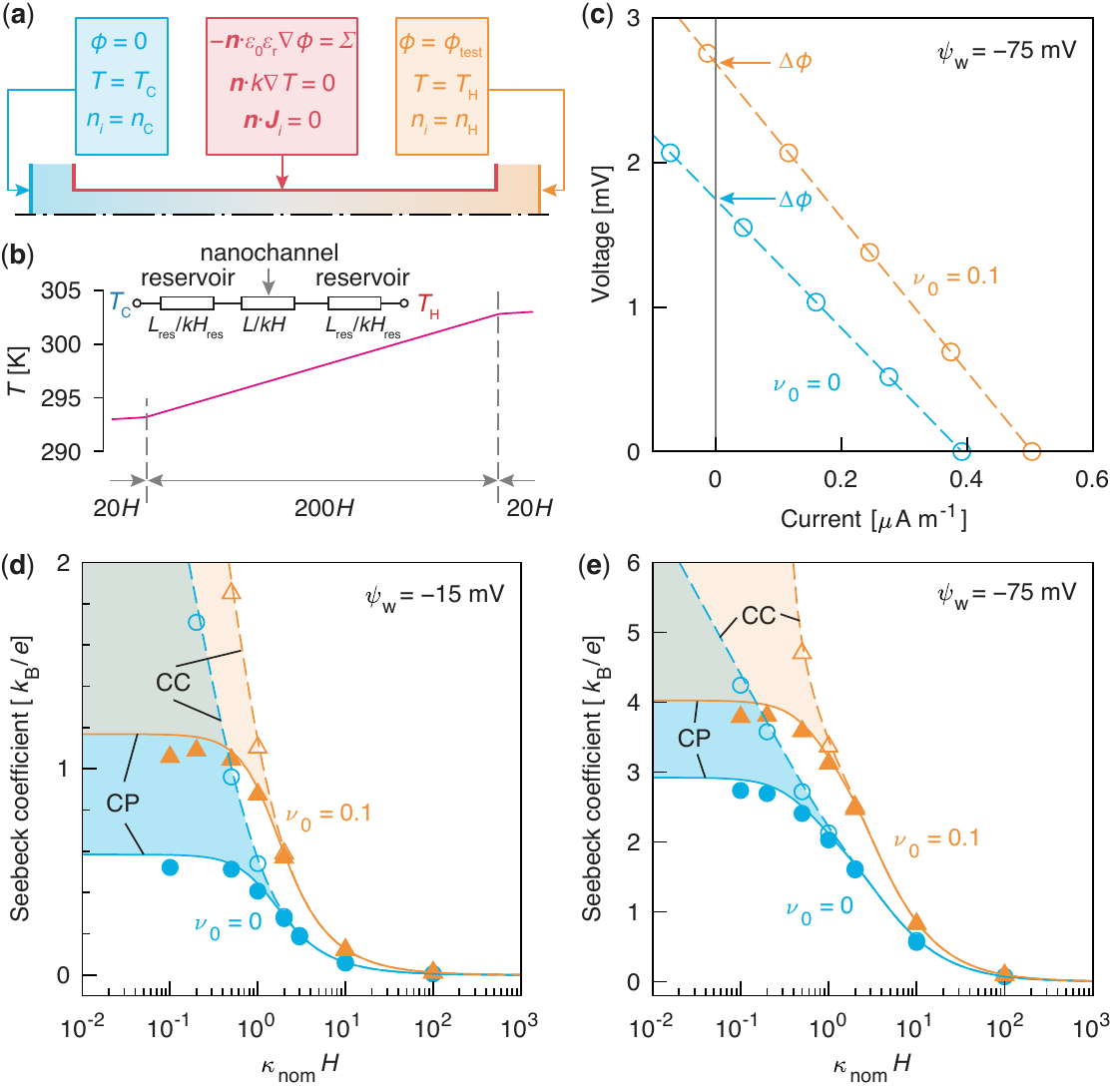}
\caption{{Simulation setup and results}. (a) Computational domain and boundary conditions. (b) Temperature distribution along the centerline of nanochannel. Inset shows the equivalent thermal resistance of the reservoir-nanochannel-reservoir system. (c) Voltage-current relationship. Symbols stand for numerical results and dash lines stand for the corresponding linear fits. (d, e) Seebeck coefficient as a function of nominal Debye parameter $\kappa_{\mathrm{nom}} H$ for CP (solid lines or solid symbols) and CC (dash lines or open symbols) boundary conditions as (d) $\psi_{\mathrm{w}}=-15$ mV and (e) $\psi_{\mathrm{w}}=-75$ mV. For convenience, here we plot the absolute values of the negative Seebeck coefficient. The lines are computed by the semi-analytical model using Eq.~(\ref{eqn:Se}) and the symbols are evaluated by the numerical simulation detailed in Supporting Information (Sec. S4). Other parameter values were the same as those in Fig.~\ref{fig:fig1}.}
\label{fig:fig4}
\end{figure*}

Then, we further confirm the new thermoelectric mechanism in electrolyte-filled nanochannels using numerical simulation. The calculation domain and boundary conditions of the simulation are shown in Fig.~\ref{fig:fig4}a and the numerical approach is detailed in Supporting Information (Sec. S4). It is worth mentioning that in the simulations, the reference concentration was kept as $n_0/N_{\rm A}=\SI{1}{mM}$ (thus $\kappa_{\rm nom}^{-1}\approx \SI{10}{nm} $) and the varying $\kappa_{\rm nom}H$ was achieved by altering $H$, here $N_{\rm A}$ is the Avogadro constant. Following the thermal resistance analysis in our previous work \cite{zhang2022sim}, we find the effective temperature difference along the nanochannel is $\sim\Delta T /(1+2 (L_{\rm res}/L)(H/H_{\rm res}) ) \approx 0.96 \Delta T $, which is very close to $\Delta T$ and confirmed by Fig.~\ref{fig:fig4}b. Therefore, the numerical and semi-analytical results are comparable. As predicted, the numerical results suggest the current-voltage ($I-V$) characteristics are linear for either $\nu_0=0$ or $\nu_0=0.1$ (Fig.~\ref{fig:fig4}c) and thus one can obtain the thermally induced potential or the open-circuit voltage (denoted by $\Delta V$) using the linear fits (see the arrows in Fig.~\ref{fig:fig4}c). Subsequently, the numerical results for the Seebeck coefficient can be evaluated as $S_{\rm e}=-\Delta V/\Delta T$. Fig.~\ref{fig:fig4}d and e indicates for either $\psi_{\rm w}=\SI{-15}{mV}$ or $\psi_{\rm w}=\SI{-75}{mV}$, the semi-analytical results coincide with the numerical results well. This agreement confirms the new thermoelectric mechanism due to the coupling of the ion steric effects and the ion thermodiffusion in the EDLs. 

We also explore the effects of the boundary condition of the Poisson-Fermi equation on the Seebeck coefficient. Usually, there are three types of boundary conditions for Poisson-Fermi equation on the charged surface, that is, the constant-potential (CP), the constant-charge (CC) and charge-regulation (CR) boundaries. The CR boundary \cite{Ritt2022}, which depends on the surface chemical reactions and frequently gives rise to results falling between CP and CC boundaries~\cite{Israelachvili2011,RuizCabello2014,Zhao2015}, is not discussed here. For given surface potential $\psi_{\rm w}$ in CP boundary, the corresponding surface charge density $\varSigma$ in CC boundary is computed from the Grahame equation. Clearly, the CP and CC boundaries result in identical $S_{\rm e}$ for $\nu_0=0$ and $\nu_0=0.1$ if there is no EDL overlap ($\kappa_{\rm nom}H\gtrsim 3$). This is because for $\nu_0=0$ or $\nu_0=0.1$, the classic or modified Grahame equations suggest the CP and CC boundaries coincide with each other in cases without EDL overlap. As $\kappa_{\rm nom}H$ decreases, the difference in $S_{\rm e}$ between CP and CC boundaries increases (see the filled areas in Fig.~\ref{fig:fig4}d, e). The more realistic prediction for the Seebeck coefficient $S_{\rm e}$ is expected to fall in the filled areas. We also notice that for the CP boundary, the semi-analytical model slightly overestimates $S_{\rm e}$ compared with the numerical simulation at small $\kappa_{\rm nom}H$. It is probably because in such cases the semi-anlytical model expressed by Eq.~(\ref{eqn:Se}) neglects the exit-entry effects which get stronger with decreasing $\kappa_{\rm nom}H$. In addition, we should emphasize that as $\kappa_{\rm nom}H\rightarrow 0$, the CC boundary may lead to an unphysically large prediction for $S_{\rm e}$.

%\subsection{Effects of short-range interionic electrostatic correlation}
%%%%%%%%%%%%%%%%%%% FIG 5 %%%%%%%%%%%%%%%%%%%
\begin{figure}[t]
\centering
\includegraphics{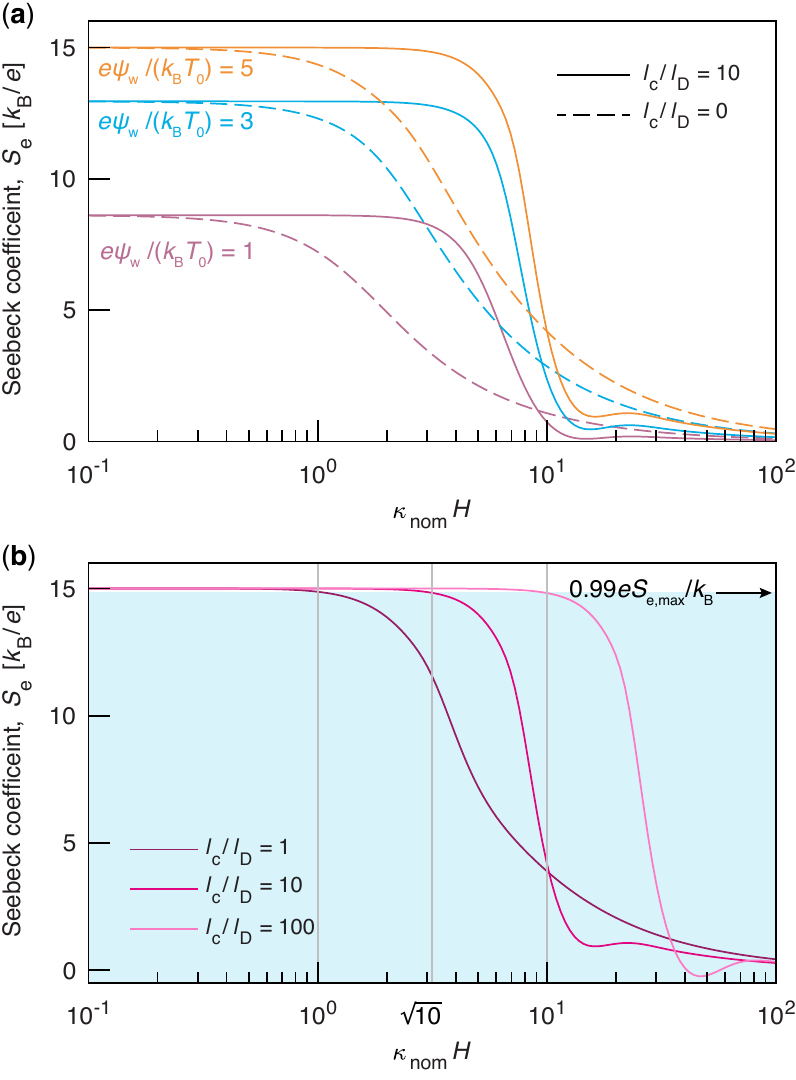}
\caption{{Effect of electrostatic correlation on thermoelectric response}. (a) Seebeck coefficient as a function of nominal Debye parameter for two ion correlation lengths of $l_{\rm c}=10l_{\rm D}=10\kappa_{\rm nom}^{-1}$ (solid lines) and $l_{\rm c}=0$ (dashed lines), for three dimensionless surface potentials of $e\psi_{\rm w}/(k_{\rm B}T_0)=1,\ 3$ and 5. (b) Seebeck coefficient as a function of nominal Debye parameter for three ion correlation lengths of $l_{\rm c}=l_{\rm D}=\kappa_{\rm nom}^{-1}$, $l_{\rm c}=10l_{\rm D}=10\kappa_{\rm nom}^{-1}$ and $l_{\rm c}=100l_{\rm D}=100\kappa_{\rm nom}^{-1}$ with $\nu_0=0.5$. The ion volume fractions $\nu_0=0.5$ and $l_{\rm c}/l_{\rm D}=1,\ 10,\ 100$ are taken from Ref.~\cite{Bazant2011}. The reduced Soret coefficients of ions are set to $\alpha_{+}=\alpha_{-}=5$ and the normalized difference in ion diffusion coefficients was set to $\chi=0$.}
\label{fig:fig5}
\end{figure}

For solvent-free ionic liquids, in addition to steric effect due to finite ion sizes, the overscreaning effect due to short-range ion correlations should be taken into account to correctly capture the EDL structure~\cite{Bazant2011}. By introducing the electrostatic correlation length, $l_{\rm c}$, Bazant et al. \cite{Bazant2011} derived a modified Poisson-Fermi equation to describe the EDL potential:
\begin{equation}\label{eq:mpf}
	\frac{\partial^2 \tilde{\psi}}{\partial y^2} - l_{\rm c}^2 \frac{\partial^4 \tilde{\psi}}{\partial y^4} \approx \frac{\kappa^2 \sinh (\tilde{\psi}) }{1-\nu  +\nu \cosh \left( \tilde{\psi} \right)}
\end{equation}
which is solved with the following boundary conditions~\cite{Bazant2011}:
\begin{equation}\label{eq:mpf_bc}
	\left.\frac{\partial^3\tilde{\psi}}{\partial{y}^3}\right|_{y=H}=0,\ \tilde{\psi}|_{y=H}=\tilde{\psi}_{\rm w},\ \left.\frac{\partial^2\tilde{\psi}}{\partial{y}^2}\right|_{y=0}=\left.\frac{\partial\tilde{\psi}}{\partial{y}}\right|_{y=0}=0
\end{equation}
It is worthwhile mentioning that the effective ion size should be modified as $({\rm \pi}/6)\sigma^3/\varPhi_{\max}$ to account for the random close packing of spheres and typically the volume fraction $\varPhi_{\max}=0.63$ \cite{Bazant2011}. We can readily find that the modification of Poisson-Fermi equation does not alter Eq.~(\ref{eqn:Se}). Equations (\ref{eq:mpf}) and (\ref{eq:mpf_bc}) are numerically solved to obtain $\tilde{\psi}$, which is further used to calculate the Seebeck coefficient using Eq.~(\ref{eqn:Se}). 

Intriguingly, the short-range electrostatic correlation does not alter the maximum values of the Seebeck coefficient that is achieved at small $\kappa_{\rm nom}H$ values, but helps to maintain such maximum values in a wider range of  $\kappa_{\rm nom}H$ values (Fig.~\ref{fig:fig5}a). Specifically, for high $e\psi_{\rm w}/(k_{\rm B}T_0)$ values the Seebeck coefficients in cases of $l_{\rm c}/l_{\rm D}=\kappa_{\rm nom}l_{\rm c}=10$ saturate to the peak values of $S_{\rm e,max}=[\psi_{\rm w}/T_0 + \tanh(e\psi_{\rm w}/k_{\rm B}T_0)(k_{\rm B}/e) \nu_0\alpha/(1-\nu_0)] $, at much larger $\kappa_{\rm nom}H$ values (as high as $\sim \sqrt{10}=\sqrt{l_{\rm c}/l_{\rm D}}=\sqrt{\kappa_{\rm nom}l_{\rm c}}$, see Fig.~\ref{fig:fig5}a and b) compared to the cases of $l_{\rm c}/l_{\rm D}=\kappa_{\rm nom}l_{\rm c}=0$ (the Seebeck coefficient saturates to peak values as $\kappa_{\rm nom}H<1$). The reason for such behavior probably lies in the decrease in the effective channel width due to the counterion crowding near the wall. The larger $\kappa_{\rm nom}H$ values imply higher ion concentrations that are beneficial to enhancing the electric conductivity~\cite{Haynes2016} and thus the thermoelctric conversion efficiency, since the thermoelectric conversion efficiency positively correlates with both the Seebeck coefficient and electric conductivity~\cite{Shi2020}.

%\section{Conclusions}

This study revealed a new thermoelectric mechanism originating from the coupling of the ion steric effect and ion thermodiffusion in the EDLs via both theoretical analysis and numerical simulation. The new mechanism can significantly enhance the Seebeck coefficient of confined electrolytes. For ionic liquids with high ion volume fractions confined by highly charged nanochannels, the enhancement relative to the cases without ion steric effect \cite{dietzel2016,Zhang2019} can reach 9.4 to 23.7 times. In addition, the short-range electrostatic correlation does not affect the value of the maximum Seebeck coefficients achieved under extreme confinement, but helps to preserve such maximum values with electrolytes of higher ion strength or weaker confinement. These findings suggest the possibility of simultaneously achieving high Seebeck coefficient and high conductivity in ionic thermoelectric conversion. This study improves the fundamental understanding of the thermoelectricity of electrolytes in nanochannels and provides theoretical guidance for related energy harvesting and temperature sensing application. In the future, (nonequilibrium) molecular dynamics simulation \cite{Fu2017,Fu2018,Fu2019,Jin2021} (Sec. S7, Supporting Information) can be carried out to provide more insight into the microscopic picture of ionic thermoelectricity in nanochannels.

%%%%%%%%%%%%%%%%%%%%%%%%%%%%% ACKNOWLEDGEMENT %%%%%%%%%%%%%%%%%%%%%%%%%%%%%%%
\textit{Acknowledgement}.---We acknowledge support by the National Natural Science Foundation of China (Nos. 51976157, 51721004). W. Z. would like to thank the support of the China Scholarship Council (No. 202106280109).
%%%%%%%%%%%%%%%%%%%%%%%%%%%%%%%%%%  REFERENCES %%%%%%%%%%%%%%%%%%%%%%%%%%%%%% 
\bibliography{mainref.bib}

%apsrev4-2.bst 2019-01-14 (MD) hand-edited version of apsrev4-1.bst
%Control: key (0)
%Control: author (8) initials jnrlst
%Control: editor formatted (1) identically to author
%Control: production of article title (0) allowed
%Control: page (0) single
%Control: year (1) truncated
%Control: production of eprint (0) enabled
\begin{thebibliography}{52}%
\makeatletter
\providecommand \@ifxundefined [1]{%
 \@ifx{#1\undefined}
}%
\providecommand \@ifnum [1]{%
 \ifnum #1\expandafter \@firstoftwo
 \else \expandafter \@secondoftwo
 \fi
}%
\providecommand \@ifx [1]{%
 \ifx #1\expandafter \@firstoftwo
 \else \expandafter \@secondoftwo
 \fi
}%
\providecommand \natexlab [1]{#1}%
\providecommand \enquote  [1]{``#1''}%
\providecommand \bibnamefont  [1]{#1}%
\providecommand \bibfnamefont [1]{#1}%
\providecommand \citenamefont [1]{#1}%
\providecommand \href@noop [0]{\@secondoftwo}%
\providecommand \href [0]{\begingroup \@sanitize@url \@href}%
\providecommand \@href[1]{\@@startlink{#1}\@@href}%
\providecommand \@@href[1]{\endgroup#1\@@endlink}%
\providecommand \@sanitize@url [0]{\catcode `\\12\catcode `\$12\catcode
  `\&12\catcode `\#12\catcode `\^12\catcode `\_12\catcode `\%12\relax}%
\providecommand \@@startlink[1]{}%
\providecommand \@@endlink[0]{}%
\providecommand \url  [0]{\begingroup\@sanitize@url \@url }%
\providecommand \@url [1]{\endgroup\@href {#1}{\urlprefix }}%
\providecommand \urlprefix  [0]{URL }%
\providecommand \Eprint [0]{\href }%
\providecommand \doibase [0]{https://doi.org/}%
\providecommand \selectlanguage [0]{\@gobble}%
\providecommand \bibinfo  [0]{\@secondoftwo}%
\providecommand \bibfield  [0]{\@secondoftwo}%
\providecommand \translation [1]{[#1]}%
\providecommand \BibitemOpen [0]{}%
\providecommand \bibitemStop [0]{}%
\providecommand \bibitemNoStop [0]{.\EOS\space}%
\providecommand \EOS [0]{\spacefactor3000\relax}%
\providecommand \BibitemShut  [1]{\csname bibitem#1\endcsname}%
\let\auto@bib@innerbib\@empty
%</preamble>
\bibitem [{\citenamefont {Bell}(2008)}]{Bell2008}%
  \BibitemOpen
  \bibfield  {author} {\bibinfo {author} {\bibfnamefont {L.~E.}\ \bibnamefont
  {Bell}},\ }\bibfield  {title} {\bibinfo {title} {Cooling, heating, generating
  power, and recovering waste heat with thermoelectric systems},\ }\href
  {https://doi.org/10.1126/science.1158899} {\bibfield  {journal} {\bibinfo
  {journal} {Science}\ }\textbf {\bibinfo {volume} {321}},\ \bibinfo {pages}
  {1457} (\bibinfo {year} {2008})}\BibitemShut {NoStop}%
\bibitem [{\citenamefont {Shi}\ \emph {et~al.}(2020)\citenamefont {Shi},
  \citenamefont {Zou},\ and\ \citenamefont {Chen}}]{Shi2020}%
  \BibitemOpen
  \bibfield  {author} {\bibinfo {author} {\bibfnamefont {X.-L.}\ \bibnamefont
  {Shi}}, \bibinfo {author} {\bibfnamefont {J.}~\bibnamefont {Zou}},\ and\
  \bibinfo {author} {\bibfnamefont {Z.-G.}\ \bibnamefont {Chen}},\ }\bibfield
  {title} {\bibinfo {title} {Advanced thermoelectric design: from materials and
  structures to devices},\ }\href {https://doi.org/10.1021/acs.chemrev.0c00026}
  {\bibfield  {journal} {\bibinfo  {journal} {Chem. Rev.}\ }\textbf {\bibinfo
  {volume} {120}},\ \bibinfo {pages} {7399} (\bibinfo {year}
  {2020})}\BibitemShut {NoStop}%
\bibitem [{\citenamefont {Herwaarden}\ and\ \citenamefont
  {Sarro}(1986)}]{Herwaarden1986}%
  \BibitemOpen
  \bibfield  {author} {\bibinfo {author} {\bibfnamefont {A.~V.}\ \bibnamefont
  {Herwaarden}}\ and\ \bibinfo {author} {\bibfnamefont {P.}~\bibnamefont
  {Sarro}},\ }\bibfield  {title} {\bibinfo {title} {Thermal sensors based on
  the seebeck effect},\ }\href {https://doi.org/10.1016/0250-6874(86)80053-1}
  {\bibfield  {journal} {\bibinfo  {journal} {Sensor. Actuator.}\ }\textbf
  {\bibinfo {volume} {10}},\ \bibinfo {pages} {321} (\bibinfo {year}
  {1986})}\BibitemShut {NoStop}%
\bibitem [{\citenamefont {Ichinose}\ \emph {et~al.}(2019)\citenamefont
  {Ichinose}, \citenamefont {Yoshida}, \citenamefont {Horiuchi}, \citenamefont
  {Fukuhara}, \citenamefont {Komatsu}, \citenamefont {Gao}, \citenamefont
  {Yomogida}, \citenamefont {Matsubara}, \citenamefont {Yamamoto},
  \citenamefont {Kono},\ and\ \citenamefont {Yanagi}}]{Ichinose2019}%
  \BibitemOpen
  \bibfield  {author} {\bibinfo {author} {\bibfnamefont {Y.}~\bibnamefont
  {Ichinose}}, \bibinfo {author} {\bibfnamefont {A.}~\bibnamefont {Yoshida}},
  \bibinfo {author} {\bibfnamefont {K.}~\bibnamefont {Horiuchi}}, \bibinfo
  {author} {\bibfnamefont {K.}~\bibnamefont {Fukuhara}}, \bibinfo {author}
  {\bibfnamefont {N.}~\bibnamefont {Komatsu}}, \bibinfo {author} {\bibfnamefont
  {W.}~\bibnamefont {Gao}}, \bibinfo {author} {\bibfnamefont {Y.}~\bibnamefont
  {Yomogida}}, \bibinfo {author} {\bibfnamefont {M.}~\bibnamefont {Matsubara}},
  \bibinfo {author} {\bibfnamefont {T.}~\bibnamefont {Yamamoto}}, \bibinfo
  {author} {\bibfnamefont {J.}~\bibnamefont {Kono}},\ and\ \bibinfo {author}
  {\bibfnamefont {K.}~\bibnamefont {Yanagi}},\ }\bibfield  {title} {\bibinfo
  {title} {Solving the thermoelectric trade-off problem with metallic carbon
  nanotubes},\ }\href {https://doi.org/10.1021/acs.nanolett.9b03022} {\bibfield
   {journal} {\bibinfo  {journal} {Nano Lett.}\ }\textbf {\bibinfo {volume}
  {19}},\ \bibinfo {pages} {7370} (\bibinfo {year} {2019})},\ \Eprint
  {https://arxiv.org/abs/https://doi.org/10.1021/acs.nanolett.9b03022}
  {https://doi.org/10.1021/acs.nanolett.9b03022} \BibitemShut {NoStop}%
\bibitem [{\citenamefont {Hurtado-Gallego}\ \emph {et~al.}(2022)\citenamefont
  {Hurtado-Gallego}, \citenamefont {Sangtarash}, \citenamefont {Davidson},
  \citenamefont {Rincón-García}, \citenamefont {Daaoub}, \citenamefont
  {Rubio-Bollinger}, \citenamefont {Lambert}, \citenamefont {Oganesyan},
  \citenamefont {Bryce}, \citenamefont {Agraït},\ and\ \citenamefont
  {Sadeghi}}]{HurtadoGallego2022}%
  \BibitemOpen
  \bibfield  {author} {\bibinfo {author} {\bibfnamefont {J.}~\bibnamefont
  {Hurtado-Gallego}}, \bibinfo {author} {\bibfnamefont {S.}~\bibnamefont
  {Sangtarash}}, \bibinfo {author} {\bibfnamefont {R.}~\bibnamefont
  {Davidson}}, \bibinfo {author} {\bibfnamefont {L.}~\bibnamefont
  {Rincón-García}}, \bibinfo {author} {\bibfnamefont {A.}~\bibnamefont
  {Daaoub}}, \bibinfo {author} {\bibfnamefont {G.}~\bibnamefont
  {Rubio-Bollinger}}, \bibinfo {author} {\bibfnamefont {C.~J.}\ \bibnamefont
  {Lambert}}, \bibinfo {author} {\bibfnamefont {V.~S.}\ \bibnamefont
  {Oganesyan}}, \bibinfo {author} {\bibfnamefont {M.~R.}\ \bibnamefont
  {Bryce}}, \bibinfo {author} {\bibfnamefont {N.}~\bibnamefont {Agraït}},\
  and\ \bibinfo {author} {\bibfnamefont {H.}~\bibnamefont {Sadeghi}},\
  }\bibfield  {title} {\bibinfo {title} {Thermoelectric enhancement in single
  organic radical molecules},\ }\href
  {https://doi.org/10.1021/acs.nanolett.1c03698} {\bibfield  {journal}
  {\bibinfo  {journal} {Nano Lett.}\ }\textbf {\bibinfo {volume} {22}},\
  \bibinfo {pages} {948} (\bibinfo {year} {2022})},\ \bibinfo {note} {pMID:
  35073099},\ \Eprint
  {https://arxiv.org/abs/https://doi.org/10.1021/acs.nanolett.1c03698}
  {https://doi.org/10.1021/acs.nanolett.1c03698} \BibitemShut {NoStop}%
\bibitem [{\citenamefont {Forman}\ \emph {et~al.}(2016)\citenamefont {Forman},
  \citenamefont {Muritala}, \citenamefont {Pardemann},\ and\ \citenamefont
  {Meyer}}]{Forman2016}%
  \BibitemOpen
  \bibfield  {author} {\bibinfo {author} {\bibfnamefont {C.}~\bibnamefont
  {Forman}}, \bibinfo {author} {\bibfnamefont {I.~K.}\ \bibnamefont
  {Muritala}}, \bibinfo {author} {\bibfnamefont {R.}~\bibnamefont
  {Pardemann}},\ and\ \bibinfo {author} {\bibfnamefont {B.}~\bibnamefont
  {Meyer}},\ }\bibfield  {title} {\bibinfo {title} {Estimating the global waste
  heat potential},\ }\href {https://doi.org/10.1016/j.rser.2015.12.192}
  {\bibfield  {journal} {\bibinfo  {journal} {Renew. Sust. Energ. Rev.}\
  }\textbf {\bibinfo {volume} {57}},\ \bibinfo {pages} {1568} (\bibinfo {year}
  {2016})}\BibitemShut {NoStop}%
\bibitem [{\citenamefont {Venkatasubramanian}(2019)}]{Venkatasubramanian2019}%
  \BibitemOpen
  \bibfield  {author} {\bibinfo {author} {\bibfnamefont {R.}~\bibnamefont
  {Venkatasubramanian}},\ }\bibfield  {title} {\bibinfo {title} {Power from
  nano-engineered wood},\ }\href {https://doi.org/10.1038/s41563-019-0352-1}
  {\bibfield  {journal} {\bibinfo  {journal} {Nat. Mater.}\ }\textbf {\bibinfo
  {volume} {18}},\ \bibinfo {pages} {536} (\bibinfo {year} {2019})}\BibitemShut
  {NoStop}%
\bibitem [{\citenamefont {Yu}\ \emph {et~al.}(2020)\citenamefont {Yu},
  \citenamefont {Duan}, \citenamefont {Cong}, \citenamefont {Xie},
  \citenamefont {Liu}, \citenamefont {Zhuang}, \citenamefont {Wang},
  \citenamefont {Qi}, \citenamefont {Xu}, \citenamefont {Wang},\ and\
  \citenamefont {Zhou}}]{Yu2020}%
  \BibitemOpen
  \bibfield  {author} {\bibinfo {author} {\bibfnamefont {B.}~\bibnamefont
  {Yu}}, \bibinfo {author} {\bibfnamefont {J.}~\bibnamefont {Duan}}, \bibinfo
  {author} {\bibfnamefont {H.}~\bibnamefont {Cong}}, \bibinfo {author}
  {\bibfnamefont {W.}~\bibnamefont {Xie}}, \bibinfo {author} {\bibfnamefont
  {R.}~\bibnamefont {Liu}}, \bibinfo {author} {\bibfnamefont {X.}~\bibnamefont
  {Zhuang}}, \bibinfo {author} {\bibfnamefont {H.}~\bibnamefont {Wang}},
  \bibinfo {author} {\bibfnamefont {B.}~\bibnamefont {Qi}}, \bibinfo {author}
  {\bibfnamefont {M.}~\bibnamefont {Xu}}, \bibinfo {author} {\bibfnamefont
  {Z.~L.}\ \bibnamefont {Wang}},\ and\ \bibinfo {author} {\bibfnamefont
  {J.}~\bibnamefont {Zhou}},\ }\bibfield  {title} {\bibinfo {title}
  {Thermosensitive crystallization{\textendash}boosted liquid thermocells for
  low-grade heat harvesting},\ }\href {https://doi.org/10.1126/science.abd6749}
  {\bibfield  {journal} {\bibinfo  {journal} {Science}\ }\textbf {\bibinfo
  {volume} {370}},\ \bibinfo {pages} {342} (\bibinfo {year}
  {2020})}\BibitemShut {NoStop}%
\bibitem [{\citenamefont {Dietzel}\ and\ \citenamefont
  {Hardt}(2016)}]{dietzel2016}%
  \BibitemOpen
  \bibfield  {author} {\bibinfo {author} {\bibfnamefont {M.}~\bibnamefont
  {Dietzel}}\ and\ \bibinfo {author} {\bibfnamefont {S.}~\bibnamefont
  {Hardt}},\ }\bibfield  {title} {\bibinfo {title} {Thermoelectricity in
  confined liquid electrolytes},\ }\href
  {https://doi.org/10.1103/PhysRevLett.116.225901} {\bibfield  {journal}
  {\bibinfo  {journal} {Phys. Rev. Lett.}\ }\textbf {\bibinfo {volume} {116}},\
  \bibinfo {pages} {225901} (\bibinfo {year} {2016})}\BibitemShut {NoStop}%
\bibitem [{\citenamefont {Li}\ \emph {et~al.}(2019)\citenamefont {Li},
  \citenamefont {Zhang}, \citenamefont {Lacey}, \citenamefont {Mi},
  \citenamefont {Zhao}, \citenamefont {Jiang}, \citenamefont {Song},
  \citenamefont {Liu}, \citenamefont {Chen}, \citenamefont {Dai}, \citenamefont
  {Yao}, \citenamefont {Das}, \citenamefont {Yang}, \citenamefont {Briber},\
  and\ \citenamefont {Hu}}]{Li2019}%
  \BibitemOpen
  \bibfield  {author} {\bibinfo {author} {\bibfnamefont {T.}~\bibnamefont
  {Li}}, \bibinfo {author} {\bibfnamefont {X.}~\bibnamefont {Zhang}}, \bibinfo
  {author} {\bibfnamefont {S.~D.}\ \bibnamefont {Lacey}}, \bibinfo {author}
  {\bibfnamefont {R.}~\bibnamefont {Mi}}, \bibinfo {author} {\bibfnamefont
  {X.}~\bibnamefont {Zhao}}, \bibinfo {author} {\bibfnamefont {F.}~\bibnamefont
  {Jiang}}, \bibinfo {author} {\bibfnamefont {J.}~\bibnamefont {Song}},
  \bibinfo {author} {\bibfnamefont {Z.}~\bibnamefont {Liu}}, \bibinfo {author}
  {\bibfnamefont {G.}~\bibnamefont {Chen}}, \bibinfo {author} {\bibfnamefont
  {J.}~\bibnamefont {Dai}}, \bibinfo {author} {\bibfnamefont {Y.}~\bibnamefont
  {Yao}}, \bibinfo {author} {\bibfnamefont {S.}~\bibnamefont {Das}}, \bibinfo
  {author} {\bibfnamefont {R.}~\bibnamefont {Yang}}, \bibinfo {author}
  {\bibfnamefont {R.~M.}\ \bibnamefont {Briber}},\ and\ \bibinfo {author}
  {\bibfnamefont {L.}~\bibnamefont {Hu}},\ }\bibfield  {title} {\bibinfo
  {title} {Cellulose ionic conductors with high differential thermal voltage
  for low-grade heat harvesting},\ }\href
  {https://doi.org/10.1038/s41563-019-0315-6} {\bibfield  {journal} {\bibinfo
  {journal} {Nat. Mater.}\ }\textbf {\bibinfo {volume} {18}},\ \bibinfo {pages}
  {608} (\bibinfo {year} {2019})}\BibitemShut {NoStop}%
\bibitem [{\citenamefont {Fu}\ \emph {et~al.}(2019)\citenamefont {Fu},
  \citenamefont {Joly},\ and\ \citenamefont {Merabia}}]{Fu2019}%
  \BibitemOpen
  \bibfield  {author} {\bibinfo {author} {\bibfnamefont {L.}~\bibnamefont
  {Fu}}, \bibinfo {author} {\bibfnamefont {L.}~\bibnamefont {Joly}},\ and\
  \bibinfo {author} {\bibfnamefont {S.}~\bibnamefont {Merabia}},\ }\bibfield
  {title} {\bibinfo {title} {Giant thermoelectric response of nanofluidic
  systems driven by water excess enthalpy},\ }\href
  {https://doi.org/10.1103/physrevlett.123.138001} {\bibfield  {journal}
  {\bibinfo  {journal} {Phys. Rev. Lett.}\ }\textbf {\bibinfo {volume} {123}},\
  \bibinfo {pages} {138001} (\bibinfo {year} {2019})}\BibitemShut {NoStop}%
\bibitem [{\citenamefont {W{\"u}rger}(2010)}]{Wurger_2010}%
  \BibitemOpen
  \bibfield  {author} {\bibinfo {author} {\bibfnamefont {A.}~\bibnamefont
  {W{\"u}rger}},\ }\bibfield  {title} {\bibinfo {title} {Thermal
  non-equilibrium transport in colloids},\ }\href
  {https://doi.org/10.1088/0034-4885/73/12/126601} {\bibfield  {journal}
  {\bibinfo  {journal} {Rep. Prog. Phys.}\ }\textbf {\bibinfo {volume} {73}},\
  \bibinfo {pages} {126601} (\bibinfo {year} {2010})}\BibitemShut {NoStop}%
\bibitem [{\citenamefont {Bonetti}\ \emph {et~al.}(2011)\citenamefont
  {Bonetti}, \citenamefont {Nakamae}, \citenamefont {Roger},\ and\
  \citenamefont {Guenoun}}]{Bonetti2011}%
  \BibitemOpen
  \bibfield  {author} {\bibinfo {author} {\bibfnamefont {M.}~\bibnamefont
  {Bonetti}}, \bibinfo {author} {\bibfnamefont {S.}~\bibnamefont {Nakamae}},
  \bibinfo {author} {\bibfnamefont {M.}~\bibnamefont {Roger}},\ and\ \bibinfo
  {author} {\bibfnamefont {P.}~\bibnamefont {Guenoun}},\ }\bibfield  {title}
  {\bibinfo {title} {Huge seebeck coefficients in nonaqueous electrolytes},\
  }\href {https://doi.org/10.1063/1.3561735} {\bibfield  {journal} {\bibinfo
  {journal} {J. Chem. Phys.}\ }\textbf {\bibinfo {volume} {134}},\ \bibinfo
  {pages} {114513} (\bibinfo {year} {2011})}\BibitemShut {NoStop}%
\bibitem [{\citenamefont {Bonetti}\ \emph {et~al.}(2015)\citenamefont
  {Bonetti}, \citenamefont {Nakamae}, \citenamefont {Huang}, \citenamefont
  {Salez}, \citenamefont {Wiertel-Gasquet},\ and\ \citenamefont
  {Roger}}]{Bonetti2015}%
  \BibitemOpen
  \bibfield  {author} {\bibinfo {author} {\bibfnamefont {M.}~\bibnamefont
  {Bonetti}}, \bibinfo {author} {\bibfnamefont {S.}~\bibnamefont {Nakamae}},
  \bibinfo {author} {\bibfnamefont {B.~T.}\ \bibnamefont {Huang}}, \bibinfo
  {author} {\bibfnamefont {T.~J.}\ \bibnamefont {Salez}}, \bibinfo {author}
  {\bibfnamefont {C.}~\bibnamefont {Wiertel-Gasquet}},\ and\ \bibinfo {author}
  {\bibfnamefont {M.}~\bibnamefont {Roger}},\ }\bibfield  {title} {\bibinfo
  {title} {Thermoelectric energy recovery at ionic-liquid/electrode
  interface},\ }\href {https://doi.org/10.1063/1.4923199} {\bibfield  {journal}
  {\bibinfo  {journal} {J. Chem. Phys.}\ }\textbf {\bibinfo {volume} {142}},\
  \bibinfo {pages} {244708} (\bibinfo {year} {2015})}\BibitemShut {NoStop}%
\bibitem [{\citenamefont {Zhao}\ \emph {et~al.}(2016)\citenamefont {Zhao},
  \citenamefont {Wang}, \citenamefont {Khan}, \citenamefont {Chen},
  \citenamefont {Gabrielsson}, \citenamefont {Jonsson}, \citenamefont
  {Berggren},\ and\ \citenamefont {Crispin}}]{Zhao2016}%
  \BibitemOpen
  \bibfield  {author} {\bibinfo {author} {\bibfnamefont {D.}~\bibnamefont
  {Zhao}}, \bibinfo {author} {\bibfnamefont {H.}~\bibnamefont {Wang}}, \bibinfo
  {author} {\bibfnamefont {Z.~U.}\ \bibnamefont {Khan}}, \bibinfo {author}
  {\bibfnamefont {J.~C.}\ \bibnamefont {Chen}}, \bibinfo {author}
  {\bibfnamefont {R.}~\bibnamefont {Gabrielsson}}, \bibinfo {author}
  {\bibfnamefont {M.~P.}\ \bibnamefont {Jonsson}}, \bibinfo {author}
  {\bibfnamefont {M.}~\bibnamefont {Berggren}},\ and\ \bibinfo {author}
  {\bibfnamefont {X.}~\bibnamefont {Crispin}},\ }\bibfield  {title} {\bibinfo
  {title} {Ionic thermoelectric supercapacitors},\ }\href
  {https://doi.org/10.1039/c6ee00121a} {\bibfield  {journal} {\bibinfo
  {journal} {Energy Environ. Sci.}\ }\textbf {\bibinfo {volume} {9}},\ \bibinfo
  {pages} {1450} (\bibinfo {year} {2016})}\BibitemShut {NoStop}%
\bibitem [{\citenamefont {Dietzel}\ and\ \citenamefont
  {Hardt}(2017)}]{Dietzel2017}%
  \BibitemOpen
  \bibfield  {author} {\bibinfo {author} {\bibfnamefont {M.}~\bibnamefont
  {Dietzel}}\ and\ \bibinfo {author} {\bibfnamefont {S.}~\bibnamefont
  {Hardt}},\ }\bibfield  {title} {\bibinfo {title} {Flow and streaming
  potential of an electrolyte in a channel with an axial temperature
  gradient},\ }\href {https://doi.org/10.1017/jfm.2016.844} {\bibfield
  {journal} {\bibinfo  {journal} {J. Fluid Mech.}\ }\textbf {\bibinfo {volume}
  {813}},\ \bibinfo {pages} {1060} (\bibinfo {year} {2017})}\BibitemShut
  {NoStop}%
\bibitem [{\citenamefont {Zhang}\ \emph {et~al.}(2019)\citenamefont {Zhang},
  \citenamefont {Wang}, \citenamefont {Zeng},\ and\ \citenamefont
  {Zhao}}]{Zhang2019}%
  \BibitemOpen
  \bibfield  {author} {\bibinfo {author} {\bibfnamefont {W.}~\bibnamefont
  {Zhang}}, \bibinfo {author} {\bibfnamefont {Q.}~\bibnamefont {Wang}},
  \bibinfo {author} {\bibfnamefont {M.}~\bibnamefont {Zeng}},\ and\ \bibinfo
  {author} {\bibfnamefont {C.}~\bibnamefont {Zhao}},\ }\bibfield  {title}
  {\bibinfo {title} {Thermoelectric effect and temperature-gradient-driven
  electrokinetic flow of electrolyte solutions in charged nanocapillaries},\
  }\href
  {https://doi.org/https://doi.org/10.1016/j.ijheatmasstransfer.2019.118569}
  {\bibfield  {journal} {\bibinfo  {journal} {Int. J. Heat Mass Transfer}\
  }\textbf {\bibinfo {volume} {143}},\ \bibinfo {pages} {118569} (\bibinfo
  {year} {2019})}\BibitemShut {NoStop}%
\bibitem [{\citenamefont {Zhong}\ and\ \citenamefont
  {Huang}(2020{\natexlab{a}})}]{Zhong2020}%
  \BibitemOpen
  \bibfield  {author} {\bibinfo {author} {\bibfnamefont {J.}~\bibnamefont
  {Zhong}}\ and\ \bibinfo {author} {\bibfnamefont {C.}~\bibnamefont {Huang}},\
  }\bibfield  {title} {\bibinfo {title} {Thermal-driven ion transport in porous
  materials for thermoelectricity applications},\ }\href
  {https://doi.org/10.1021/acs.langmuir.9b03141} {\bibfield  {journal}
  {\bibinfo  {journal} {Langmuir}\ }\textbf {\bibinfo {volume} {36}},\ \bibinfo
  {pages} {1418} (\bibinfo {year} {2020}{\natexlab{a}})}\BibitemShut {NoStop}%
\bibitem [{\citenamefont {Zhong}\ and\ \citenamefont
  {Huang}(2020{\natexlab{b}})}]{Zhong2020a}%
  \BibitemOpen
  \bibfield  {author} {\bibinfo {author} {\bibfnamefont {J.}~\bibnamefont
  {Zhong}}\ and\ \bibinfo {author} {\bibfnamefont {C.}~\bibnamefont {Huang}},\
  }\bibfield  {title} {\bibinfo {title} {Influence factors of thermal driven
  ion transport in nano-channel for thermoelectricity application},\ }\href
  {https://doi.org/10.1016/j.ijheatmasstransfer.2020.119501} {\bibfield
  {journal} {\bibinfo  {journal} {Int. J. Heat Mass Transfer}\ }\textbf
  {\bibinfo {volume} {152}},\ \bibinfo {pages} {119501} (\bibinfo {year}
  {2020}{\natexlab{b}})}\BibitemShut {NoStop}%
\bibitem [{\citenamefont {Zhang}\ \emph {et~al.}(2022)\citenamefont {Zhang},
  \citenamefont {Farhan}, \citenamefont {Jiao}, \citenamefont {Qian},
  \citenamefont {Guo}, \citenamefont {Wang}, \citenamefont {Yang},\ and\
  \citenamefont {Zhao}}]{zhang2022sim}%
  \BibitemOpen
  \bibfield  {author} {\bibinfo {author} {\bibfnamefont {W.}~\bibnamefont
  {Zhang}}, \bibinfo {author} {\bibfnamefont {M.}~\bibnamefont {Farhan}},
  \bibinfo {author} {\bibfnamefont {K.}~\bibnamefont {Jiao}}, \bibinfo {author}
  {\bibfnamefont {F.}~\bibnamefont {Qian}}, \bibinfo {author} {\bibfnamefont
  {P.}~\bibnamefont {Guo}}, \bibinfo {author} {\bibfnamefont {Q.}~\bibnamefont
  {Wang}}, \bibinfo {author} {\bibfnamefont {C.~C.}\ \bibnamefont {Yang}},\
  and\ \bibinfo {author} {\bibfnamefont {C.}~\bibnamefont {Zhao}},\ }\bibfield
  {title} {\bibinfo {title} {Simultaneous thermoosmotic and thermoelectric
  responses in nanoconfined electrolyte solutions: Effects of nanopore
  structures and membrane properties},\ }\href
  {https://doi.org/https://doi.org/10.1016/j.jcis.2022.03.079} {\bibfield
  {journal} {\bibinfo  {journal} {J. Colloid Interface Sci.}\ }\textbf
  {\bibinfo {volume} {618}},\ \bibinfo {pages} {333} (\bibinfo {year}
  {2022})}\BibitemShut {NoStop}%
\bibitem [{\citenamefont {Qian}\ \emph {et~al.}(2022)\citenamefont {Qian},
  \citenamefont {Liu},\ and\ \citenamefont {Yang}}]{Qian2022confinement}%
  \BibitemOpen
  \bibfield  {author} {\bibinfo {author} {\bibfnamefont {X.}~\bibnamefont
  {Qian}}, \bibinfo {author} {\bibfnamefont {T.-H.}\ \bibnamefont {Liu}},\ and\
  \bibinfo {author} {\bibfnamefont {R.}~\bibnamefont {Yang}},\ }\bibfield
  {title} {\bibinfo {title} {Confinement effect on thermopower of
  electrolytes},\ }\href
  {https://doi.org/https://doi.org/10.1016/j.mtphys.2022.100627} {\bibfield
  {journal} {\bibinfo  {journal} {Mater. Today Phys.}\ }\textbf {\bibinfo
  {volume} {23}},\ \bibinfo {pages} {100627} (\bibinfo {year}
  {2022})}\BibitemShut {NoStop}%
\bibitem [{\citenamefont {Schoch}\ \emph {et~al.}(2008)\citenamefont {Schoch},
  \citenamefont {Han},\ and\ \citenamefont {Renaud}}]{Schoch2008}%
  \BibitemOpen
  \bibfield  {author} {\bibinfo {author} {\bibfnamefont {R.~B.}\ \bibnamefont
  {Schoch}}, \bibinfo {author} {\bibfnamefont {J.}~\bibnamefont {Han}},\ and\
  \bibinfo {author} {\bibfnamefont {P.}~\bibnamefont {Renaud}},\ }\bibfield
  {title} {\bibinfo {title} {Transport phenomena in nanofluidics},\ }\href
  {https://doi.org/10.1103/revmodphys.80.839} {\bibfield  {journal} {\bibinfo
  {journal} {Rev. Mod. Phys.}\ }\textbf {\bibinfo {volume} {80}},\ \bibinfo
  {pages} {839} (\bibinfo {year} {2008})}\BibitemShut {NoStop}%
\bibitem [{\citenamefont {Bocquet}\ and\ \citenamefont
  {Charlaix}(2010)}]{Bocquet2010}%
  \BibitemOpen
  \bibfield  {author} {\bibinfo {author} {\bibfnamefont {L.}~\bibnamefont
  {Bocquet}}\ and\ \bibinfo {author} {\bibfnamefont {E.}~\bibnamefont
  {Charlaix}},\ }\bibfield  {title} {\bibinfo {title} {Nanofluidics, from bulk
  to interfaces},\ }\href {https://doi.org/10.1039/b909366b} {\bibfield
  {journal} {\bibinfo  {journal} {Chem. Soc. Rev.}\ }\textbf {\bibinfo {volume}
  {39}},\ \bibinfo {pages} {1073} (\bibinfo {year} {2010})}\BibitemShut
  {NoStop}%
\bibitem [{\citenamefont {Borukhov}\ \emph {et~al.}(1997)\citenamefont
  {Borukhov}, \citenamefont {Andelman},\ and\ \citenamefont
  {Orland}}]{Borukhov1997}%
  \BibitemOpen
  \bibfield  {author} {\bibinfo {author} {\bibfnamefont {I.}~\bibnamefont
  {Borukhov}}, \bibinfo {author} {\bibfnamefont {D.}~\bibnamefont {Andelman}},\
  and\ \bibinfo {author} {\bibfnamefont {H.}~\bibnamefont {Orland}},\
  }\bibfield  {title} {\bibinfo {title} {Steric effects in electrolytes: a
  modified poisson-boltzmann equation},\ }\href
  {https://doi.org/10.1103/PhysRevLett.79.435} {\bibfield  {journal} {\bibinfo
  {journal} {Phys. Rev. Lett.}\ }\textbf {\bibinfo {volume} {79}},\ \bibinfo
  {pages} {435} (\bibinfo {year} {1997})}\BibitemShut {NoStop}%
\bibitem [{\citenamefont {Tessier}\ and\ \citenamefont
  {Slater}(2006)}]{Tessier2006}%
  \BibitemOpen
  \bibfield  {author} {\bibinfo {author} {\bibfnamefont {F.}~\bibnamefont
  {Tessier}}\ and\ \bibinfo {author} {\bibfnamefont {G.~W.}\ \bibnamefont
  {Slater}},\ }\bibfield  {title} {\bibinfo {title} {Effective debye length in
  closed nanoscopic systems: A competition between two length scales},\ }\href
  {https://doi.org/10.1002/elps.200500457} {\bibfield  {journal} {\bibinfo
  {journal} {Electrophoresis}\ }\textbf {\bibinfo {volume} {27}},\ \bibinfo
  {pages} {686} (\bibinfo {year} {2006})}\BibitemShut {NoStop}%
\bibitem [{\citenamefont {Bazant}\ \emph {et~al.}(2009)\citenamefont {Bazant},
  \citenamefont {Kilic}, \citenamefont {Storey},\ and\ \citenamefont
  {Ajdari}}]{Bazant2009}%
  \BibitemOpen
  \bibfield  {author} {\bibinfo {author} {\bibfnamefont {M.~Z.}\ \bibnamefont
  {Bazant}}, \bibinfo {author} {\bibfnamefont {M.~S.}\ \bibnamefont {Kilic}},
  \bibinfo {author} {\bibfnamefont {B.~D.}\ \bibnamefont {Storey}},\ and\
  \bibinfo {author} {\bibfnamefont {A.}~\bibnamefont {Ajdari}},\ }\bibfield
  {title} {\bibinfo {title} {Towards an understanding of induced-charge
  electrokinetics at large applied voltages in concentrated solutions},\ }\href
  {https://doi.org/10.1016/j.cis.2009.10.001} {\bibfield  {journal} {\bibinfo
  {journal} {Adv. Colloid Interface Sci.}\ }\textbf {\bibinfo {volume} {152}},\
  \bibinfo {pages} {48} (\bibinfo {year} {2009})}\BibitemShut {NoStop}%
\bibitem [{\citenamefont {Bazant}\ \emph {et~al.}(2011)\citenamefont {Bazant},
  \citenamefont {Storey},\ and\ \citenamefont {Kornyshev}}]{Bazant2011}%
  \BibitemOpen
  \bibfield  {author} {\bibinfo {author} {\bibfnamefont {M.~Z.}\ \bibnamefont
  {Bazant}}, \bibinfo {author} {\bibfnamefont {B.~D.}\ \bibnamefont {Storey}},\
  and\ \bibinfo {author} {\bibfnamefont {A.~A.}\ \bibnamefont {Kornyshev}},\
  }\bibfield  {title} {\bibinfo {title} {Double layer in ionic liquids:
  overscreening versus crowding},\ }\href
  {https://doi.org/10.1103/physrevlett.106.046102} {\bibfield  {journal}
  {\bibinfo  {journal} {Phys. Rev. Lett.}\ }\textbf {\bibinfo {volume} {106}},\
  \bibinfo {pages} {046102} (\bibinfo {year} {2011})}\BibitemShut {NoStop}%
\bibitem [{\citenamefont {Huang}\ \emph {et~al.}(2015)\citenamefont {Huang},
  \citenamefont {Roger}, \citenamefont {Bonetti}, \citenamefont {Salez},
  \citenamefont {Wiertel-Gasquet}, \citenamefont {Dubois}, \citenamefont
  {Gomes}, \citenamefont {Demouchy}, \citenamefont {M{\'{e}}riguet},
  \citenamefont {Peyre}, \citenamefont {Kouyat{\'{e}}}, \citenamefont
  {Filomeno}, \citenamefont {Depeyrot}, \citenamefont {Tourinho}, \citenamefont
  {Perzynski},\ and\ \citenamefont {Nakamae}}]{Huang2015}%
  \BibitemOpen
  \bibfield  {author} {\bibinfo {author} {\bibfnamefont {B.~T.}\ \bibnamefont
  {Huang}}, \bibinfo {author} {\bibfnamefont {M.}~\bibnamefont {Roger}},
  \bibinfo {author} {\bibfnamefont {M.}~\bibnamefont {Bonetti}}, \bibinfo
  {author} {\bibfnamefont {T.~J.}\ \bibnamefont {Salez}}, \bibinfo {author}
  {\bibfnamefont {C.}~\bibnamefont {Wiertel-Gasquet}}, \bibinfo {author}
  {\bibfnamefont {E.}~\bibnamefont {Dubois}}, \bibinfo {author} {\bibfnamefont
  {R.~C.}\ \bibnamefont {Gomes}}, \bibinfo {author} {\bibfnamefont
  {G.}~\bibnamefont {Demouchy}}, \bibinfo {author} {\bibfnamefont
  {G.}~\bibnamefont {M{\'{e}}riguet}}, \bibinfo {author} {\bibfnamefont
  {V.}~\bibnamefont {Peyre}}, \bibinfo {author} {\bibfnamefont
  {M.}~\bibnamefont {Kouyat{\'{e}}}}, \bibinfo {author} {\bibfnamefont {C.~L.}\
  \bibnamefont {Filomeno}}, \bibinfo {author} {\bibfnamefont {J.}~\bibnamefont
  {Depeyrot}}, \bibinfo {author} {\bibfnamefont {F.~A.}\ \bibnamefont
  {Tourinho}}, \bibinfo {author} {\bibfnamefont {R.}~\bibnamefont
  {Perzynski}},\ and\ \bibinfo {author} {\bibfnamefont {S.}~\bibnamefont
  {Nakamae}},\ }\bibfield  {title} {\bibinfo {title} {Thermoelectricity and
  thermodiffusion in charged colloids},\ }\href
  {https://doi.org/10.1063/1.4927665} {\bibfield  {journal} {\bibinfo
  {journal} {J. Chem. Phys.}\ }\textbf {\bibinfo {volume} {143}},\ \bibinfo
  {pages} {054902} (\bibinfo {year} {2015})}\BibitemShut {NoStop}%
\bibitem [{\citenamefont {van~der Heyden}\ \emph {et~al.}(2006)\citenamefont
  {van~der Heyden}, \citenamefont {Bonthuis}, \citenamefont {Stein},
  \citenamefont {Meyer},\ and\ \citenamefont {Dekker}}]{Heyden2006}%
  \BibitemOpen
  \bibfield  {author} {\bibinfo {author} {\bibfnamefont {F.~H.~J.}\
  \bibnamefont {van~der Heyden}}, \bibinfo {author} {\bibfnamefont {D.~J.}\
  \bibnamefont {Bonthuis}}, \bibinfo {author} {\bibfnamefont {D.}~\bibnamefont
  {Stein}}, \bibinfo {author} {\bibfnamefont {C.}~\bibnamefont {Meyer}},\ and\
  \bibinfo {author} {\bibfnamefont {C.}~\bibnamefont {Dekker}},\ }\bibfield
  {title} {\bibinfo {title} {Electrokinetic energy conversion efficiency in
  nanofluidic channels},\ }\href {https://doi.org/10.1021/nl061524l} {\bibfield
   {journal} {\bibinfo  {journal} {Nano Lett.}\ }\textbf {\bibinfo {volume}
  {6}},\ \bibinfo {pages} {2232} (\bibinfo {year} {2006})}\BibitemShut
  {NoStop}%
\bibitem [{\citenamefont {Siria}\ \emph {et~al.}(2013)\citenamefont {Siria},
  \citenamefont {Poncharal}, \citenamefont {Biance}, \citenamefont {Fulcrand},
  \citenamefont {Blase}, \citenamefont {Purcell},\ and\ \citenamefont
  {Bocquet}}]{Siria2013}%
  \BibitemOpen
  \bibfield  {author} {\bibinfo {author} {\bibfnamefont {A.}~\bibnamefont
  {Siria}}, \bibinfo {author} {\bibfnamefont {P.}~\bibnamefont {Poncharal}},
  \bibinfo {author} {\bibfnamefont {A.-L.}\ \bibnamefont {Biance}}, \bibinfo
  {author} {\bibfnamefont {R.}~\bibnamefont {Fulcrand}}, \bibinfo {author}
  {\bibfnamefont {X.}~\bibnamefont {Blase}}, \bibinfo {author} {\bibfnamefont
  {S.~T.}\ \bibnamefont {Purcell}},\ and\ \bibinfo {author} {\bibfnamefont
  {L.}~\bibnamefont {Bocquet}},\ }\bibfield  {title} {\bibinfo {title} {Giant
  osmotic energy conversion measured in a single transmembrane boron nitride
  nanotube},\ }\href {https://doi.org/10.1038/nature11876} {\bibfield
  {journal} {\bibinfo  {journal} {Nature}\ }\textbf {\bibinfo {volume} {494}},\
  \bibinfo {pages} {455} (\bibinfo {year} {2013})}\BibitemShut {NoStop}%
\bibitem [{\citenamefont {Feng}\ \emph {et~al.}(2016)\citenamefont {Feng},
  \citenamefont {Graf}, \citenamefont {Liu}, \citenamefont {Ovchinnikov},
  \citenamefont {Dumcenco}, \citenamefont {Heiranian}, \citenamefont
  {Nandigana}, \citenamefont {Aluru}, \citenamefont {Kis},\ and\ \citenamefont
  {Radenovic}}]{Feng2016}%
  \BibitemOpen
  \bibfield  {author} {\bibinfo {author} {\bibfnamefont {J.}~\bibnamefont
  {Feng}}, \bibinfo {author} {\bibfnamefont {M.}~\bibnamefont {Graf}}, \bibinfo
  {author} {\bibfnamefont {K.}~\bibnamefont {Liu}}, \bibinfo {author}
  {\bibfnamefont {D.}~\bibnamefont {Ovchinnikov}}, \bibinfo {author}
  {\bibfnamefont {D.}~\bibnamefont {Dumcenco}}, \bibinfo {author}
  {\bibfnamefont {M.}~\bibnamefont {Heiranian}}, \bibinfo {author}
  {\bibfnamefont {V.}~\bibnamefont {Nandigana}}, \bibinfo {author}
  {\bibfnamefont {N.~R.}\ \bibnamefont {Aluru}}, \bibinfo {author}
  {\bibfnamefont {A.}~\bibnamefont {Kis}},\ and\ \bibinfo {author}
  {\bibfnamefont {A.}~\bibnamefont {Radenovic}},\ }\bibfield  {title} {\bibinfo
  {title} {Single-layer \ce{MoS2} nanopores as nanopower generators},\ }\href
  {https://doi.org/10.1038/nature18593} {\bibfield  {journal} {\bibinfo
  {journal} {Nature}\ }\textbf {\bibinfo {volume} {536}},\ \bibinfo {pages}
  {197} (\bibinfo {year} {2016})}\BibitemShut {NoStop}%
\bibitem [{\citenamefont {Zhang}\ \emph {et~al.}(2021)\citenamefont {Zhang},
  \citenamefont {Wen},\ and\ \citenamefont {Jiang}}]{Zhang2021}%
  \BibitemOpen
  \bibfield  {author} {\bibinfo {author} {\bibfnamefont {Z.}~\bibnamefont
  {Zhang}}, \bibinfo {author} {\bibfnamefont {L.}~\bibnamefont {Wen}},\ and\
  \bibinfo {author} {\bibfnamefont {L.}~\bibnamefont {Jiang}},\ }\bibfield
  {title} {\bibinfo {title} {Nanofluidics for osmotic energy conversion},\
  }\href {https://doi.org/10.1038/s41578-021-00300-4} {\bibfield  {journal}
  {\bibinfo  {journal} {Nat. Rev. Mater.}\ }\textbf {\bibinfo {volume} {6}},\
  \bibinfo {pages} {622} (\bibinfo {year} {2021})}\BibitemShut {NoStop}%
\bibitem [{\citenamefont {Ritt}\ \emph {et~al.}(2022)\citenamefont {Ritt},
  \citenamefont {de~Souza}, \citenamefont {Barsukov}, \citenamefont {Yosinski},
  \citenamefont {Bazant}, \citenamefont {Reed},\ and\ \citenamefont
  {Elimelech}}]{Ritt2022}%
  \BibitemOpen
  \bibfield  {author} {\bibinfo {author} {\bibfnamefont {C.~L.}\ \bibnamefont
  {Ritt}}, \bibinfo {author} {\bibfnamefont {J.~P.}\ \bibnamefont {de~Souza}},
  \bibinfo {author} {\bibfnamefont {M.~G.}\ \bibnamefont {Barsukov}}, \bibinfo
  {author} {\bibfnamefont {S.}~\bibnamefont {Yosinski}}, \bibinfo {author}
  {\bibfnamefont {M.~Z.}\ \bibnamefont {Bazant}}, \bibinfo {author}
  {\bibfnamefont {M.~A.}\ \bibnamefont {Reed}},\ and\ \bibinfo {author}
  {\bibfnamefont {M.}~\bibnamefont {Elimelech}},\ }\bibfield  {title} {\bibinfo
  {title} {Thermodynamics of charge regulation during ion transport through
  silica nanochannels},\ }\href {https://doi.org/10.1021/acsnano.2c06633}
  {\bibfield  {journal} {\bibinfo  {journal} {{ACS} Nano}\ }\textbf {\bibinfo
  {volume} {16}},\ \bibinfo {pages} {15249} (\bibinfo {year}
  {2022})}\BibitemShut {NoStop}%
\bibitem [{\citenamefont {W\"urger}(2020)}]{wurger2020thermopower}%
  \BibitemOpen
  \bibfield  {author} {\bibinfo {author} {\bibfnamefont {A.}~\bibnamefont
  {W\"urger}},\ }\bibfield  {title} {\bibinfo {title} {Thermopower of ionic
  conductors and ionic capacitors},\ }\href
  {https://doi.org/10.1103/PhysRevResearch.2.042030} {\bibfield  {journal}
  {\bibinfo  {journal} {Phys. Rev. Res.}\ }\textbf {\bibinfo {volume} {2}},\
  \bibinfo {pages} {042030} (\bibinfo {year} {2020})}\BibitemShut {NoStop}%
\bibitem [{\citenamefont {Fair}\ and\ \citenamefont
  {Osterle}(1971)}]{Fair1971}%
  \BibitemOpen
  \bibfield  {author} {\bibinfo {author} {\bibfnamefont {J.~C.}\ \bibnamefont
  {Fair}}\ and\ \bibinfo {author} {\bibfnamefont {J.~F.}\ \bibnamefont
  {Osterle}},\ }\bibfield  {title} {\bibinfo {title} {Reverse electrodialysis
  in charged capillary membranes},\ }\href {https://doi.org/10.1063/1.1675344}
  {\bibfield  {journal} {\bibinfo  {journal} {J. Chem. Phys.}\ }\textbf
  {\bibinfo {volume} {54}},\ \bibinfo {pages} {3307} (\bibinfo {year}
  {1971})}\BibitemShut {NoStop}%
\bibitem [{\citenamefont {Peters}\ \emph {et~al.}(2016)\citenamefont {Peters},
  \citenamefont {van Roij}, \citenamefont {Bazant},\ and\ \citenamefont
  {Biesheuvel}}]{Peters2016}%
  \BibitemOpen
  \bibfield  {author} {\bibinfo {author} {\bibfnamefont {P.~B.}\ \bibnamefont
  {Peters}}, \bibinfo {author} {\bibfnamefont {R.}~\bibnamefont {van Roij}},
  \bibinfo {author} {\bibfnamefont {M.~Z.}\ \bibnamefont {Bazant}},\ and\
  \bibinfo {author} {\bibfnamefont {P.~M.}\ \bibnamefont {Biesheuvel}},\
  }\bibfield  {title} {\bibinfo {title} {Analysis of electrolyte transport
  through charged nanopores},\ }\href
  {https://doi.org/10.1103/PhysRevE.93.053108} {\bibfield  {journal} {\bibinfo
  {journal} {Phys. Rev. E}\ }\textbf {\bibinfo {volume} {93}},\ \bibinfo
  {pages} {053108} (\bibinfo {year} {2016})}\BibitemShut {NoStop}%
\bibitem [{\citenamefont {Alizadeh}\ and\ \citenamefont
  {Mani}(2017)}]{Alizadeh2017}%
  \BibitemOpen
  \bibfield  {author} {\bibinfo {author} {\bibfnamefont {S.}~\bibnamefont
  {Alizadeh}}\ and\ \bibinfo {author} {\bibfnamefont {A.}~\bibnamefont
  {Mani}},\ }\bibfield  {title} {\bibinfo {title} {Multiscale model for
  electrokinetic transport in networks of pores, part i: Model derivation},\
  }\href {https://doi.org/10.1021/acs.langmuir.6b03816} {\bibfield  {journal}
  {\bibinfo  {journal} {Langmuir}\ }\textbf {\bibinfo {volume} {33}},\ \bibinfo
  {pages} {6205} (\bibinfo {year} {2017})}\BibitemShut {NoStop}%
\bibitem [{\citenamefont {Baldessari}(2008)}]{Baldessari2008}%
  \BibitemOpen
  \bibfield  {author} {\bibinfo {author} {\bibfnamefont {F.}~\bibnamefont
  {Baldessari}},\ }\bibfield  {title} {\bibinfo {title} {Electrokinetics in
  nanochannels: Part i. electric double layer overlap and channel-to-well
  equilibrium},\ }\href
  {https://doi.org/https://doi.org/10.1016/j.jcis.2008.06.007} {\bibfield
  {journal} {\bibinfo  {journal} {J. Colloid Interface Sci.}\ }\textbf
  {\bibinfo {volume} {325}},\ \bibinfo {pages} {526} (\bibinfo {year}
  {2008})}\BibitemShut {NoStop}%
\bibitem [{\citenamefont {Volkov}\ \emph {et~al.}(1997)\citenamefont {Volkov},
  \citenamefont {Paula},\ and\ \citenamefont {Deamer}}]{Volkov1997}%
  \BibitemOpen
  \bibfield  {author} {\bibinfo {author} {\bibfnamefont {A.}~\bibnamefont
  {Volkov}}, \bibinfo {author} {\bibfnamefont {S.}~\bibnamefont {Paula}},\ and\
  \bibinfo {author} {\bibfnamefont {D.}~\bibnamefont {Deamer}},\ }\bibfield
  {title} {\bibinfo {title} {Two mechanisms of permeation of small neutral
  molecules and hydrated ions across phospholipid bilayers},\ }\href
  {https://doi.org/10.1016/S0302-4598(96)05097-0} {\bibfield  {journal}
  {\bibinfo  {journal} {Bioelectrochem. Bioenerg.}\ }\textbf {\bibinfo {volume}
  {42}},\ \bibinfo {pages} {153} (\bibinfo {year} {1997})}\BibitemShut
  {NoStop}%
\bibitem [{\citenamefont {Fedorov}\ and\ \citenamefont
  {Kornyshev}(2014)}]{Fedorov2014}%
  \BibitemOpen
  \bibfield  {author} {\bibinfo {author} {\bibfnamefont {M.~V.}\ \bibnamefont
  {Fedorov}}\ and\ \bibinfo {author} {\bibfnamefont {A.~A.}\ \bibnamefont
  {Kornyshev}},\ }\bibfield  {title} {\bibinfo {title} {Ionic liquids at
  electrified interfaces},\ }\href {https://doi.org/10.1021/cr400374x}
  {\bibfield  {journal} {\bibinfo  {journal} {Chem. Rev.}\ }\textbf {\bibinfo
  {volume} {114}},\ \bibinfo {pages} {2978} (\bibinfo {year}
  {2014})}\BibitemShut {NoStop}%
\bibitem [{\citenamefont {Gebbie}\ \emph {et~al.}(2015)\citenamefont {Gebbie},
  \citenamefont {Dobbs}, \citenamefont {Valtiner},\ and\ \citenamefont
  {Israelachvili}}]{Gebbie2015}%
  \BibitemOpen
  \bibfield  {author} {\bibinfo {author} {\bibfnamefont {M.~A.}\ \bibnamefont
  {Gebbie}}, \bibinfo {author} {\bibfnamefont {H.~A.}\ \bibnamefont {Dobbs}},
  \bibinfo {author} {\bibfnamefont {M.}~\bibnamefont {Valtiner}},\ and\
  \bibinfo {author} {\bibfnamefont {J.~N.}\ \bibnamefont {Israelachvili}},\
  }\bibfield  {title} {\bibinfo {title} {Long-range electrostatic screening in
  ionic liquids},\ }\href {https://doi.org/10.1073/pnas.1508366112} {\bibfield
  {journal} {\bibinfo  {journal} {Proc. Natl. Acad. Sci.}\ }\textbf {\bibinfo
  {volume} {112}},\ \bibinfo {pages} {7432} (\bibinfo {year}
  {2015})}\BibitemShut {NoStop}%
\bibitem [{\citenamefont {Hatlo}\ \emph {et~al.}(2012)\citenamefont {Hatlo},
  \citenamefont {van Roij},\ and\ \citenamefont {Lue}}]{Hatlo2012}%
  \BibitemOpen
  \bibfield  {author} {\bibinfo {author} {\bibfnamefont {M.~M.}\ \bibnamefont
  {Hatlo}}, \bibinfo {author} {\bibfnamefont {R.}~\bibnamefont {van Roij}},\
  and\ \bibinfo {author} {\bibfnamefont {L.}~\bibnamefont {Lue}},\ }\bibfield
  {title} {\bibinfo {title} {The electric double layer at high surface
  potentials: The influence of excess ion polarizability},\ }\href
  {https://doi.org/10.1209/0295-5075/97/28010} {\bibfield  {journal} {\bibinfo
  {journal} {Europhys. Lett.}\ }\textbf {\bibinfo {volume} {97}},\ \bibinfo
  {pages} {28010} (\bibinfo {year} {2012})}\BibitemShut {NoStop}%
\bibitem [{\citenamefont {Biesheuvel}(2011)}]{Biesheuvel2011}%
  \BibitemOpen
  \bibfield  {author} {\bibinfo {author} {\bibfnamefont {P.}~\bibnamefont
  {Biesheuvel}},\ }\bibfield  {title} {\bibinfo {title} {Two-fluid model for
  the simultaneous flow of colloids and fluids in porous media},\ }\href
  {https://doi.org/10.1016/j.jcis.2010.12.006} {\bibfield  {journal} {\bibinfo
  {journal} {J. Colloid Interface Sci.}\ }\textbf {\bibinfo {volume} {355}},\
  \bibinfo {pages} {389} (\bibinfo {year} {2011})}\BibitemShut {NoStop}%
\bibitem [{\citenamefont {Würger}(2021)}]{Wuerger2021}%
  \BibitemOpen
  \bibfield  {author} {\bibinfo {author} {\bibfnamefont {A.}~\bibnamefont
  {Würger}},\ }\bibfield  {title} {\bibinfo {title} {Thermoelectric ratchet
  effect for charge carriers with hopping dynamics},\ }\href
  {https://doi.org/10.1103/physrevlett.126.068001} {\bibfield  {journal}
  {\bibinfo  {journal} {Phys. Rev. Lett.}\ }\textbf {\bibinfo {volume} {126}},\
  \bibinfo {pages} {068001} (\bibinfo {year} {2021})}\BibitemShut {NoStop}%
\bibitem [{\citenamefont {D'Agostino}\ \emph {et~al.}(2018)\citenamefont
  {D'Agostino}, \citenamefont {Mantle}, \citenamefont {Mullan}, \citenamefont
  {Hardacre},\ and\ \citenamefont {Gladden}}]{DAgostino2018}%
  \BibitemOpen
  \bibfield  {author} {\bibinfo {author} {\bibfnamefont {C.}~\bibnamefont
  {D'Agostino}}, \bibinfo {author} {\bibfnamefont {M.~D.}\ \bibnamefont
  {Mantle}}, \bibinfo {author} {\bibfnamefont {C.~L.}\ \bibnamefont {Mullan}},
  \bibinfo {author} {\bibfnamefont {C.}~\bibnamefont {Hardacre}},\ and\
  \bibinfo {author} {\bibfnamefont {L.~F.}\ \bibnamefont {Gladden}},\
  }\bibfield  {title} {\bibinfo {title} {Ddiffusion, ion pairing and
  aggregation in 1-ethyl-3-methylimidazolium-based ionic liquids studied by
  \ce{^1H} and \ce{^19F} \ce{PFG} \ce{NMR}: Effect of temperature, anion and
  glucose dissolution},\ }\href {https://doi.org/10.1002/cphc.201701354}
  {\bibfield  {journal} {\bibinfo  {journal} {ChemPhysChem}\ }\textbf {\bibinfo
  {volume} {19}},\ \bibinfo {pages} {1081} (\bibinfo {year}
  {2018})}\BibitemShut {NoStop}%
\bibitem [{\citenamefont {Israelachvili}(2011)}]{Israelachvili2011}%
  \BibitemOpen
  \bibfield  {author} {\bibinfo {author} {\bibfnamefont {J.~N.}\ \bibnamefont
  {Israelachvili}},\ }\href@noop {} {\emph {\bibinfo {title} {Intermolecular
  and Surface Forces}}},\ \bibinfo {edition} {3rd}\ ed.\ (\bibinfo  {publisher}
  {Academic Press},\ \bibinfo {address} {San Diego},\ \bibinfo {year}
  {2011})\BibitemShut {NoStop}%
\bibitem [{\citenamefont {Ruiz-Cabello}\ \emph {et~al.}(2014)\citenamefont
  {Ruiz-Cabello}, \citenamefont {Trefalt}, \citenamefont {Maroni},\ and\
  \citenamefont {Borkovec}}]{RuizCabello2014}%
  \BibitemOpen
  \bibfield  {author} {\bibinfo {author} {\bibfnamefont {F.~J.~M.}\
  \bibnamefont {Ruiz-Cabello}}, \bibinfo {author} {\bibfnamefont
  {G.}~\bibnamefont {Trefalt}}, \bibinfo {author} {\bibfnamefont
  {P.}~\bibnamefont {Maroni}},\ and\ \bibinfo {author} {\bibfnamefont
  {M.}~\bibnamefont {Borkovec}},\ }\bibfield  {title} {\bibinfo {title}
  {Electric double-layer potentials and surface regulation properties measured
  by colloidal-probe atomic force microscopy},\ }\href
  {https://doi.org/10.1103/physreve.90.012301} {\bibfield  {journal} {\bibinfo
  {journal} {Phys. Rev. E}\ }\textbf {\bibinfo {volume} {90}},\ \bibinfo
  {pages} {012301} (\bibinfo {year} {2014})}\BibitemShut {NoStop}%
\bibitem [{\citenamefont {Zhao}\ \emph {et~al.}(2015)\citenamefont {Zhao},
  \citenamefont {Ebeling}, \citenamefont {Siretanu}, \citenamefont {van~den
  Ende},\ and\ \citenamefont {Mugele}}]{Zhao2015}%
  \BibitemOpen
  \bibfield  {author} {\bibinfo {author} {\bibfnamefont {C.}~\bibnamefont
  {Zhao}}, \bibinfo {author} {\bibfnamefont {D.}~\bibnamefont {Ebeling}},
  \bibinfo {author} {\bibfnamefont {I.}~\bibnamefont {Siretanu}}, \bibinfo
  {author} {\bibfnamefont {D.}~\bibnamefont {van~den Ende}},\ and\ \bibinfo
  {author} {\bibfnamefont {F.}~\bibnamefont {Mugele}},\ }\bibfield  {title}
  {\bibinfo {title} {Extracting local surface charges and charge regulation
  behavior from atomic force microscopy measurements at heterogeneous
  solid-electrolyte interfaces},\ }\href {https://doi.org/10.1039/C5NR05261K}
  {\bibfield  {journal} {\bibinfo  {journal} {Nanoscale}\ }\textbf {\bibinfo
  {volume} {7}},\ \bibinfo {pages} {16298} (\bibinfo {year}
  {2015})}\BibitemShut {NoStop}%
\bibitem [{\citenamefont {Haynes}\ \emph {et~al.}(2016)\citenamefont {Haynes},
  \citenamefont {Lide},\ and\ \citenamefont {Bruno}}]{Haynes2016}%
  \BibitemOpen
  \bibfield  {author} {\bibinfo {author} {\bibfnamefont {W.~M.}\ \bibnamefont
  {Haynes}}, \bibinfo {author} {\bibfnamefont {D.~R.}\ \bibnamefont {Lide}},\
  and\ \bibinfo {author} {\bibfnamefont {T.~J.}\ \bibnamefont {Bruno}},\
  }\href@noop {} {\emph {\bibinfo {title} {CRC Handbook of Chemistry and
  Physics}}},\ \bibinfo {edition} {97th}\ ed.\ (\bibinfo  {publisher} {CRC
  press},\ \bibinfo {address} {Boca Raton},\ \bibinfo {year}
  {2016})\BibitemShut {NoStop}%
\bibitem [{\citenamefont {Fu}\ \emph {et~al.}(2017)\citenamefont {Fu},
  \citenamefont {Merabia},\ and\ \citenamefont {Joly}}]{Fu2017}%
  \BibitemOpen
  \bibfield  {author} {\bibinfo {author} {\bibfnamefont {L.}~\bibnamefont
  {Fu}}, \bibinfo {author} {\bibfnamefont {S.}~\bibnamefont {Merabia}},\ and\
  \bibinfo {author} {\bibfnamefont {L.}~\bibnamefont {Joly}},\ }\bibfield
  {title} {\bibinfo {title} {What controls thermo-osmosis? molecular
  simulations show the critical role of interfacial hydrodynamics},\ }\href
  {https://doi.org/10.1103/PhysRevLett.119.214501} {\bibfield  {journal}
  {\bibinfo  {journal} {Phys. Rev. Lett.}\ }\textbf {\bibinfo {volume} {119}},\
  \bibinfo {pages} {214501} (\bibinfo {year} {2017})}\BibitemShut {NoStop}%
\bibitem [{\citenamefont {Fu}\ \emph {et~al.}(2018)\citenamefont {Fu},
  \citenamefont {Merabia},\ and\ \citenamefont {Joly}}]{Fu2018}%
  \BibitemOpen
  \bibfield  {author} {\bibinfo {author} {\bibfnamefont {L.}~\bibnamefont
  {Fu}}, \bibinfo {author} {\bibfnamefont {S.}~\bibnamefont {Merabia}},\ and\
  \bibinfo {author} {\bibfnamefont {L.}~\bibnamefont {Joly}},\ }\bibfield
  {title} {\bibinfo {title} {Understanding fast and robust thermo-osmotic flows
  through carbon nanotube membranes: Thermodynamics meets hydrodynamics},\
  }\href {https://doi.org/10.1021/acs.jpclett.8b00703} {\bibfield  {journal}
  {\bibinfo  {journal} {J. Phys. Chem. Lett.}\ }\textbf {\bibinfo {volume}
  {9}},\ \bibinfo {pages} {2086} (\bibinfo {year} {2018})}\BibitemShut
  {NoStop}%
\bibitem [{\citenamefont {Jin}\ \emph {et~al.}(2021)\citenamefont {Jin},
  \citenamefont {Tao}, \citenamefont {Luo},\ and\ \citenamefont
  {Li}}]{Jin2021}%
  \BibitemOpen
  \bibfield  {author} {\bibinfo {author} {\bibfnamefont {Y.}~\bibnamefont
  {Jin}}, \bibinfo {author} {\bibfnamefont {R.}~\bibnamefont {Tao}}, \bibinfo
  {author} {\bibfnamefont {S.}~\bibnamefont {Luo}},\ and\ \bibinfo {author}
  {\bibfnamefont {Z.}~\bibnamefont {Li}},\ }\bibfield  {title} {\bibinfo
  {title} {Size-sensitive thermoelectric properties of electrolyte-based
  nanofluidic systems},\ }\href {https://doi.org/10.1021/acs.jpclett.0c03558}
  {\bibfield  {journal} {\bibinfo  {journal} {J. Phys. Chem. Lett.}\ }\textbf
  {\bibinfo {volume} {12}},\ \bibinfo {pages} {1144} (\bibinfo {year}
  {2021})}\BibitemShut {NoStop}%
\end{thebibliography}%

\clearpage


\section*{Supplementary material}

# Supplementary materials *for*

# Ion steric effect induces giant enhancement of thermoelectric conversion in electrolyte-filled nanochannels

Wenyao Zhang,[1] Xinxi Liu,[1] Kai Jiao,[1] Qiuwang Wang,[1] Chun Yang,[2] and Cunlu Zhao[1, *]

[1] *MOE Key Laboratory of Thermo-Fluid Science and Engineering,*

*School of Energy and Power Engineering,*

*Xi'an Jiaotong University, Xi'an 710049, China*

[2] *School of Mechanical and Aerospace Engineering, Nanyang Technological University,*

*50 Nanyang Avenue, Singapore 639798, Singapore*

## CONTENTS

* mclzhao@xjtu.edu.cn

## S1.  DERIVATION OF COUPLED THERMAL-IONIC TRANSPORT EQUATIONS

For the coupled thermal-ionic transport, the ionic flux $\boldsymbol{J}_i$ and heat flux $\boldsymbol{J}_q$ can be described by the phenomenological Onsager theory as [1, 2]

$$\boldsymbol{J}_i = L_{ii}\left(-\frac{\nabla\tilde{\mu}_i}{T}\right) + L_{iq}\nabla\left(\frac{1}{T}\right) \tag{S1}$$

$$\boldsymbol{J}_q = \sum_i L_{qi}\left(-\frac{\nabla\tilde{\mu}_i}{T}\right) + L_{qq}\nabla\left(\frac{1}{T}\right) \tag{S2}$$

where $\tilde{\mu}_i = \mu_i + z_i e\phi$ with $\mu_i$ being the chemical potential of ionic species $i$, $z_i$ the valance of ionic species $i$, $\phi$ the electrostatic potential, $e$ the elementary charge; $T$ is the absolute temperature; $L_{ii}$, $L_{iq}$, $L_{qi}$ and $L_{qq}$ are the phenomenological coefficients. Note that $\mu_i = \mu_i(\{n_i\}, T)$, we have

$$\nabla\tilde{\mu}_i = \nabla^{(T)}\tilde{\mu}_i + \left(\frac{\partial\mu_i}{\partial T}\right)_{n_i}\nabla T = \nabla^{(T)}\mu_i + z_i e\nabla\phi + \left(\frac{\partial\mu_i}{\partial T}\right)_{n_i}\nabla T \tag{S3}$$

with $\nabla^{(T)}\mu_i$ being given as [3]

$$\nabla^{(T)}\mu_i = \frac{k_{\mathrm{B}}T}{n_i}\nabla n_i + \frac{\sigma^3 k_{\mathrm{B}}T}{1-\sigma^3\sum_j n_j}\sum_i \nabla n_i \tag{S4}$$

and $(\partial_T\mu_i)_{n_i} = -s_i$ [4], where $s_i$ is the partial entropy of ionic species $i$. Substituting (S3) and (S4) into (S1) yields

$$\boldsymbol{J}_i = -\frac{L_{ii}k_{\mathrm{B}}}{n_i}\left(\nabla n_i + \frac{\sigma^3 n_i\sum_j\nabla n_j}{1-\sigma^3\sum_j n_j} + \frac{z_i en_i}{k_{\mathrm{B}}T}\nabla\phi + \frac{L_{iq}/L_{ii} - s_iT}{k_{\mathrm{B}}T^2}n_i\nabla T\right) \tag{S5}$$

Defining the ionic diffusion coefficients $D_i$ by $D_i = L_{ii}k_{\mathrm{B}}/n_i$ and the transported heats $Q_i^*$ by $Q_i^* = L_{iq}/L_{ii}$, one finds

$$\boldsymbol{J}_i = -D_i\left(\nabla n_i + \frac{\sigma^3 n_i\sum_j\nabla n_j}{1-\sigma^3\sum_j n_j} + \frac{z_i en_i}{k_{\mathrm{B}}T}\nabla\phi + \frac{2n_i\alpha_i}{T}\nabla T\right) \tag{S6}$$

where $\alpha_i = (Q_i^* - s_iT)/(2k_{\mathrm{B}}T)$ is the reduced Soret coefficient [2]. Equation (S6) is the so-called modified Nernst-Planck equation.

## S2.  ELECTRIC DOUBLE LAYER AND "FERMI-LIKE" ION DISTRIBUTIONS

The mass conservation of the ions in the investigated system leads to

$$\nabla\cdot\boldsymbol{J}_i = 0 \tag{S7}$$

Substituting (S6) into (S7) yields

$$\nabla \cdot \left[ D_i \left( \nabla n_i + \frac{\sigma^3 n_i \sum_j \nabla n_j}{1 - \sigma^3 \sum_j n_j} + \frac{z_i e n_i}{k_B T} \nabla \phi + \frac{2 n_i \alpha_i}{T} \nabla T \right) \right] = 0 \tag{S8}$$

To facilitate analysis, we introduce dimensionless variables for the axial coordinate, the lateral coordinate, the ion number concentration, the electric potential and the temperature, respectively, as

$$\tilde{x} = \frac{x}{L}, \quad \tilde{y} = \frac{y}{H}, \quad \tilde{n}_i = \frac{n_i}{\frac{1}{2} \sum_i z_i^2 n_{i,0}} \equiv \frac{n_i}{\Gamma_0}, \quad \tilde{\phi} = \frac{\phi}{k_B T_0/e} \equiv \frac{\phi}{V_{\mathrm{T}}}, \quad \tilde{T} = \frac{T}{T_0} \tag{S9}$$

where $\Gamma_0 = \frac{1}{2} \sum_i z_i^2 n_{i,0} = n_0$ is the ionic strength in the bulk solution with $n_{i,0}$ being the bulk number concentration of either ionic species, $V_{\mathrm{T}} = k_B T_0/e$ is the thermal voltage defined at room temperature $T_0$.

Accordingly, we find that (S8) can be rewritten in dimensionless form as

$$\delta^2 \frac{\partial}{\partial \tilde{x}} \left[ \frac{D_i}{D_0} \left( \frac{\partial \tilde{n}_i}{\partial \tilde{x}} + \frac{(\nu_0/2) \tilde{n}_i}{1 - (\nu_0/2) \sum_j \tilde{n}_j} \sum_j \frac{\partial \tilde{n}_j}{\partial \tilde{x}} + \frac{z_i \tilde{n}_i}{\tilde{T}} \frac{\partial \tilde{\phi}}{\partial \tilde{x}} + \frac{2 \tilde{n}_i \alpha_i}{\tilde{T}} \frac{\partial \tilde{T}}{\partial \tilde{x}} \right) \right]$$
$$+ \frac{\partial}{\partial \tilde{y}} \left[ \frac{D_i}{D_0} \left( \frac{\partial \tilde{n}_i}{\partial \tilde{y}} + \frac{(\nu_0/2) \tilde{n}_i}{1 - (\nu_0/2) \sum_j \tilde{n}_j} \sum_j \frac{\partial \tilde{n}_j}{\partial \tilde{y}} + \frac{z_i \tilde{n}_i}{\tilde{T}} \frac{\partial \tilde{\phi}}{\partial \tilde{y}} + \frac{2 \tilde{n}_i \alpha_i}{\tilde{T}} \frac{\partial \tilde{T}}{\partial \tilde{y}} \right) \right] = 0 \tag{S10}$$

where $\delta = H/L$ is the aspect ratio of the nanochannel, $\nu_0 = \sigma^3 n_0$ is the reference volume fraction of dissolved ions [5] and $D_0$ is the reference ion diffusion coefficient. Under the conditions of $\delta^2 \ll 1$, neglecting all terms of $O(\delta^2)$ and higher, (S10) can be reduced to

$$\frac{\partial}{\partial \tilde{y}} \left[ \frac{D_i}{D_0} \left( \frac{\partial \tilde{n}_i}{\partial \tilde{y}} + \frac{(\nu_0/2) \tilde{n}_i}{1 - (\nu_0/2) \sum_j \tilde{n}_j} \sum_j \frac{\partial \tilde{n}_j}{\partial \tilde{y}} + \frac{z_i \tilde{n}_i}{\tilde{T}} \frac{\partial \tilde{\phi}}{\partial \tilde{y}} + \frac{2 \tilde{n}_i \alpha_i}{\tilde{T}} \frac{\partial \tilde{T}}{\partial \tilde{y}} \right) \right] = 0 \tag{S11}$$

Integrating (S11) together with the symmetry boundary at centerline of the nanochannel and noting that $\partial \tilde{T}/\partial \tilde{y} = 0$ (since $T$ is a linear function of $x$ only [6]), we find

$$\frac{\partial \tilde{n}_i}{\partial \tilde{y}} + \frac{(\nu_0/2) \tilde{n}_i}{1 - (\nu_0/2) \sum_j \tilde{n}_j} \sum_j \frac{\partial \tilde{n}_j}{\partial \tilde{y}} + \frac{z_i \tilde{n}_i}{\tilde{T}} \frac{\partial \tilde{\phi}}{\partial \tilde{y}} = 0 \tag{S12}$$

or

$$\frac{\partial}{\partial \tilde{y}} \ln \frac{\tilde{n}_i}{1 - (\nu_0/2) \sum_j \tilde{n}_j} + \frac{z_i}{\tilde{T}} \frac{\partial \tilde{\phi}}{\partial \tilde{y}} = 0 \tag{S13}$$

Integrating (S13) in $y$ direction yields

$$\frac{\tilde{n}_i(\tilde{x}, \tilde{y})}{1 - (\nu_0/2) \sum_j \tilde{n}_j(\tilde{x}, \tilde{y})} = \frac{\tilde{n}_{i,\mathrm{v}}(\tilde{x})}{1 - (\nu_0/2) \sum_j \tilde{n}_{j,\mathrm{v}}(\tilde{x})} \exp \left[ -\frac{z_i \tilde{\phi}(\tilde{x}, \tilde{y})}{\tilde{T}(\tilde{x})} \right] \tag{S14}$$

where $\tilde{n}_{i,\mathrm{v}}(\tilde{x})$ is the (dimensionless) concentration of ionic species $i$ at virtual electroneutral solution [7, 8]. According to the literature results [7, 8], the overall potential $\tilde{\phi}(\tilde{x}, \tilde{y})$ can be divided into two parts: one is the EDL potential $\tilde{\psi}(\tilde{x}, \tilde{y})$ and the other is the induced potential or the virtual potential $\tilde{\phi}_{\mathrm{v}}(\tilde{x})$. For a symmetric $z : z$ electrolyte, it is more convenient to express the ion distributions in terms of the same $x$-dependent virtual concentration. To this end, we made a product of (S14) for cations and that for anions:

$$\frac{\sqrt{\tilde{n}_+(\tilde{x}, \tilde{y})\tilde{n}_-(\tilde{x})}}{1 - (\nu/2)[\tilde{n}_+(\tilde{x}, \tilde{y}) + \tilde{n}_-(\tilde{x}, \tilde{y})]} = \frac{\sqrt{\tilde{n}_{+,\mathrm{v}}(\tilde{x})\tilde{n}_{-,\mathrm{v}}(\tilde{x})}}{1 - (\nu/2)[\tilde{n}_{+,\mathrm{v}}(\tilde{x}) + \tilde{n}_{-,\mathrm{v}}(\tilde{x})]} = \frac{\tilde{n}_{\mathrm{v}}(\tilde{x})}{1 - \nu\tilde{n}_{\mathrm{v}}(\tilde{x})} \tag{S15}$$

which defines the same (dimensionless) virtual concentration $\tilde{n}_{\mathrm{v}}$. In addition, $\nu = 2\sigma^3 n_{\mathrm{v}}$ is the volume fraction of dissolved ions in local viratual reservoir.

Taking advantage of (S14) and (S15), we find that the virtual potential can be defined as

$$\exp\left[\frac{z\tilde{\phi}_{\mathrm{v}}(\tilde{x})}{\tilde{T}(\tilde{x})}\right] = \sqrt{\frac{\tilde{n}_{+,\mathrm{v}}(\tilde{x})}{\tilde{n}_{-,\mathrm{v}}(\tilde{x})}} = \sqrt{\frac{\tilde{n}_+(\tilde{x}, \tilde{y})}{\tilde{n}_-(\tilde{x}, \tilde{y})}} \exp\left[\frac{z\tilde{\phi}(\tilde{x}, \tilde{y})}{\tilde{T}(\tilde{x})}\right] \tag{S16}$$

Evidently, (S15) and (S16) recover the literature expression [8] as $\tilde{T} \to 1$ (in the absense of temperature gradient) and $\nu \to 0$. Substituting (S15) and (S16) into (S14) leads to

$$\frac{\tilde{n}_i(\tilde{x}, \tilde{y})}{1 - (\nu/2)\sum_j \tilde{n}_j(\tilde{x}, \tilde{y})} = \frac{\tilde{n}_{\mathrm{v}}(\tilde{x})}{1 - \nu\tilde{n}_{\mathrm{v}}(\tilde{x})} \exp\left[-\frac{z_i\tilde{\phi}(\tilde{x}, \tilde{y})}{\tilde{T}(\tilde{x})}\right] \tag{S17}$$

Subsequently, we can solve (S17) to obtain the ion distribution of either ionic species:

$$\tilde{n}_i(\tilde{x}, \tilde{y}) = \frac{\tilde{n}_{\mathrm{v}}(\tilde{x}) \exp\left[-\frac{z_i\tilde{\psi}(\tilde{x}, \tilde{y})}{\tilde{T}(\tilde{x})}\right]}{1 - \nu\tilde{n}_{\mathrm{v}}(\tilde{x}) + \nu\tilde{n}_{\mathrm{v}}(\tilde{x})\cosh\left[\frac{z\tilde{\psi}(\tilde{x}, \tilde{y})}{\tilde{T}(\tilde{x})}\right]} \tag{S18}$$

which can be rewritten in dimensional form as

$$n_i(x, y) = \frac{n_{\mathrm{v}} \exp\left(-\frac{z_i e\psi}{k_{\mathrm{B}}T}\right)}{1 - 2\sigma^3 n_{\mathrm{v}} + 2\sigma^3 n_{\mathrm{v}}\cosh\left(\frac{ze\psi}{k_{\mathrm{B}}T}\right)} \tag{S19}$$

Equation (S19) indicates that the ions satisfy the "Fermi-like" distribution.

## S3. IONIC SEEBECK COEFFICIENT UNDER CONFINEMENT

At the steady state, the Seebeck coefficient or the thermopower of an electrolyte-filled nanochannel in a system without mass exchanging with the environment can be derived

under the condition of zero current, $I = 2\int_0^H \sum_i ez_i J_{i,x} = 0$. First of all, we should derive the expression for the electric current. To this end, we need to make some mathematical transformations of the axial ion flux. From (S6), we can readily obtain the axial component of the ion flux for a symmetry $z : z$ electrolyte (Note that $\sigma^3(n_+ + n_-) = \nu\cosh(\tilde{\psi})/[1 - \nu + \nu\cosh(\tilde{\psi})] < 1$ local volume fraction of dissolved ions satisfies $0 \leq \nu < 1$):

$$J_{\pm,x} = n_\pm D_\pm \left[ \frac{\partial}{\partial x} \ln \frac{n_\pm}{1 - \sigma^3(n_+ + n_-)} \pm \frac{ze}{k_B T}\left(\frac{\partial \psi}{\partial x} - E\right) + \frac{2\alpha_\pm}{T}\frac{\partial T}{\partial x}\right] \tag{S20}$$

Making use of (S19), we find

$$\frac{n_\pm}{1 - \sigma^3(n_+ + n_-)} = \frac{n}{1 - 2\sigma^3 n}\exp\left(\mp\frac{ze\psi}{k_B T}\right) \tag{S21}$$

For convenience, hereafter the subscript v in $n_v$ is omitted.

Substituting (S21) into (S20), with some mathematical operations, we find

$$J_{i,x} = -n_i D_i \left[\frac{1}{1 - 2\sigma^3 n}\frac{d\ln n}{dx} + \left(\frac{2\alpha_i}{T} + \frac{ez_i\psi}{k_B T^2}\right)\frac{dT}{dx} - \frac{ez_i E}{k_B T}\right] \tag{S22}$$

For a symmetric $z : z$ electrolyte, the electric current is derived as

$$I = -\frac{2ez}{1 - 2\sigma^3 n}\frac{d\ln n}{dx}\int_0^H (n_+ D_+ - n_- D_-)\mathrm{d}y - \frac{4ez}{T}\frac{dT}{dx}\int_0^H (n_+ D_+\alpha_+ - n_- D_-\alpha_-)\mathrm{d}y$$

$$- \frac{2ez}{T}\frac{dT}{dx}\int_0^H (n_+ D_+ + n_- D_-)\frac{ez\psi}{k_B T}\mathrm{d}y + \frac{2e^2 z^2 E}{k_B T}\int_0^H (n_+ D_+ + n_- D_-)\mathrm{d}y. \tag{S23}$$

Substituting (S19) into (S23) yields

$$I = \frac{4eznD}{1 - 2\sigma^3 n}\frac{d\ln n}{dx}\int_0^H \frac{\sinh\left(\frac{ez\psi}{k_B T}\right) - \chi\cosh\left(\frac{ez\psi}{k_B T}\right)}{1 - 2\sigma^3 n + 2\sigma^3 n\cosh\left(\frac{ez\psi}{k_B T}\right)}\mathrm{d}y - \frac{4ezn}{T}\frac{dT}{dx}$$

$$\times \left\{D_+\delta\alpha\int_0^H \frac{\exp\left(-\frac{ez\psi}{k_B T}\right)}{1 - 2\sigma^3 n + 2\sigma^3 n\cosh\left(\frac{ez\psi}{k_B T}\right)}\mathrm{d}y - 2D\alpha_-\int_0^H \frac{\sinh\left(\frac{ez\psi}{k_B T}\right) - \chi\cosh\left(\frac{ez\psi}{k_B T}\right)}{1 - 2\sigma^3 n + 2\sigma^3 n\cosh\left(\frac{ez\psi}{k_B T}\right)}\mathrm{d}y\right\}$$

$$- \frac{4eznD}{T}\frac{dT}{dx}\int_0^H \frac{ez\psi}{k_B T}\frac{\cosh\left(\frac{ez\psi}{k_B T}\right) - \chi\sinh\left(\frac{ez\psi}{k_B T}\right)}{1 - 2\sigma^3 n + 2\sigma^3 n\cosh\left(\frac{ez\psi}{k_B T}\right)}\mathrm{d}y$$

$$+ \frac{4e^2 z^2 nDE}{k_B T}\int_0^H \frac{\cosh\left(\frac{ez\psi}{k_B T}\right) - \chi\sinh\left(\frac{ez\psi}{k_B T}\right)}{1 - 2\sigma^3 n + 2\sigma^3 n\cosh\left(\frac{ez\psi}{k_B T}\right)}\mathrm{d}y \tag{S24}$$

where $\chi = (D_+ - D_-)/(D_+ + D_-)$, $D = (D_+ + D_-)/2$ and $\delta\alpha = \alpha_+ - \alpha_-$.

Here, we assume that $n$ satisfies the classic Soret equilibrium expressed by (1) in the main text, i.e., $\mathrm{d}_x \ln n = -(\alpha/T)\mathrm{d}_x T$. The definition of variable $n$ (or $n_v$) depends on the

degree of EDL overlap. When ther is no EDL overlap, $n$ represents the ionic strength (or the concentration of either ion species) at the midplane of the channel (Fig. S3a) and $n$ must equal the number concentration of either ion species in the reservoir in the absence of external thermodynamic force (such as temperature gradient). This case has been validated for classic Soret equilibrium in bulk solutions (without EDL overlap) under a temperature gradient [9]. However, when the EDLs in the nanochannel, $n$ cannot be defined by the concentration of either ion species at channel midplane anymore. Instead, $n$ should be defined as the concentration of either ion species in a virtual electroneutral reservoir (where $\psi = 0$) that is in equilibrium with any cross section [8, 10] (Fig. S3b). The concept of virtual concentration still holds under nonisothermal conditions and for the system considered in this study, $n$ obeys the classic Soret equilibrium [6]. Accordingly, (S24) becomes

$$
\begin{aligned}
I = -\frac{4eznD}{T}\frac{\mathrm{d}T}{\mathrm{d}x} & \left\{ \frac{\alpha}{1-2\sigma^3 n} \int_0^H \frac{\sinh\left(\frac{ez\psi}{k_\mathrm{B}T}\right) - \chi\cosh\left(\frac{ez\psi}{k_\mathrm{B}T}\right)}{1-2\sigma^3 n + 2\sigma^3 n\cosh\left(\frac{ez\psi}{k_\mathrm{B}T}\right)}\mathrm{d}y + (1+\chi)\delta\alpha \right. \\
& \times \int_0^H \frac{\exp\left(-\frac{ez\psi}{k_\mathrm{B}T}\right)}{1-2\sigma^3 n + 2\sigma^3 n\cosh\left(\frac{ez\psi}{k_\mathrm{B}T}\right)}\mathrm{d}y - 2\alpha_- \int_0^H \frac{\sinh\left(\frac{ez\psi}{k_\mathrm{B}T}\right) - \chi\cosh\left(\frac{ez\psi}{k_\mathrm{B}T}\right)}{1-2\sigma^3 n + 2\sigma^3 n\cosh\left(\frac{ez\psi}{k_\mathrm{B}T}\right)}\mathrm{d}y \\
& \left. + \int_0^H \frac{ez\psi}{k_\mathrm{B}T}\frac{\cosh\left(\frac{ez\psi}{k_\mathrm{B}T}\right) - \chi\sinh\left(\frac{ez\psi}{k_\mathrm{B}T}\right)}{1-2\sigma^3 n + 2\sigma^3 n\cosh\left(\frac{ez\psi}{k_\mathrm{B}T}\right)}\mathrm{d}y \right\} \\
& + \frac{4e^2 z^2 nDE}{k_\mathrm{B}T} \int_0^H \frac{\cosh\left(\frac{ez\psi}{k_\mathrm{B}T}\right) - \chi\sinh\left(\frac{ez\psi}{k_\mathrm{B}T}\right)}{1-2\sigma^3 n + 2\sigma^3 n\cosh\left(\frac{ez\psi}{k_\mathrm{B}T}\right)}\mathrm{d}y
\end{aligned}
\tag{S25}
$$

Note that $2\alpha_- = \alpha - \delta\alpha$, (S25) can be reduced to

$$
\begin{aligned}
I = -\frac{4eznD}{T}\frac{\mathrm{d}T}{\mathrm{d}x} & \left\{ \frac{2\sigma^3 n\alpha}{1-2\sigma^3 n} \int_0^H \frac{\sinh\left(\frac{ez\psi}{k_\mathrm{B}T}\right) - \chi\cosh\left(\frac{ez\psi}{k_\mathrm{B}T}\right)}{1-2\sigma^3 n + 2\sigma^3 n\cosh\left(\frac{ez\psi}{k_\mathrm{B}T}\right)}\mathrm{d}y \right. \\
& \left. + \delta\alpha \int_0^H \frac{\cosh\left(\frac{ez\psi}{k_\mathrm{B}T}\right) - \chi\sinh\left(\frac{ez\psi}{k_\mathrm{B}T}\right)}{1-2\sigma^3 n + 2\sigma^3 n\cosh\left(\frac{ez\psi}{k_\mathrm{B}T}\right)}\mathrm{d}y + \int_0^H \frac{ez\psi}{k_\mathrm{B}T}\frac{\cosh\left(\frac{ez\psi}{k_\mathrm{B}T}\right) - \chi\sinh\left(\frac{ez\psi}{k_\mathrm{B}T}\right)}{1-2\sigma^3 n + 2\sigma^3 n\cosh\left(\frac{ez\psi}{k_\mathrm{B}T}\right)}\mathrm{d}y \right\} \\
& + \frac{4e^2 z^2 nDE}{k_\mathrm{B}T} \int_0^H \frac{\cosh\left(\frac{ez\psi}{k_\mathrm{B}T}\right) - \chi\sinh\left(\frac{ez\psi}{k_\mathrm{B}T}\right)}{1-2\sigma^3 n + 2\sigma^3 n\cosh\left(\frac{ez\psi}{k_\mathrm{B}T}\right)}\mathrm{d}y
\end{aligned}
\tag{S26}
$$

Setting $I = 0$, we obtain the expression for the Seebeck coefficient (or the thermopower):

$$S_e = \frac{E}{dT/dx} = \frac{k_B \delta\alpha}{ez} + \frac{k_B}{ez} \frac{\int_0^H \frac{ez\psi}{k_B T} \frac{\cosh\left(\frac{ez\psi}{k_B T}\right) - \chi \sinh\left(\frac{ez\psi}{k_B T}\right)}{1 - \nu + \nu \cosh\left(\frac{ez\psi}{k_B T}\right)} dy}{\int_0^H \frac{\cosh\left(\frac{ez\psi}{k_B T}\right) - \chi \sinh\left(\frac{ez\psi}{k_B T}\right)}{1 - \nu + \nu \cosh\left(\frac{ez\psi}{k_B T}\right)} dy}$$

$$+ \frac{k_B}{ez} \frac{\nu\alpha}{1-\nu} \frac{\int_0^H \frac{\sinh\left(\frac{ez\psi}{k_B T}\right) - \chi \cosh\left(\frac{ez\psi}{k_B T}\right)}{1 - \nu + \nu \cosh\left(\frac{ez\psi}{k_B T}\right)} dy}{\int_0^H \frac{\cosh\left(\frac{ez\psi}{k_B T}\right) - \chi \sinh\left(\frac{ez\psi}{k_B T}\right)}{1 - \nu + \nu \cosh\left(\frac{ez\psi}{k_B T}\right)} dy} \tag{S27}$$

which is Eq. (5) in the main text.

When the ion steric effect is negligible (i.e., $\nu \to 0$), (S27) can be simplified to the theoretical model given by Dietzel and Hardt [6]. When there is no nanoconfinement (i.e., $H \gg \kappa_{nom}^{-1}$ and thus $\psi \to 0$), (S27) can reduce to

$$S_e^{(H \gg \kappa_{nom}^{-1})} = \frac{k_B \delta\alpha}{ez} - \chi \frac{k_B}{ez} \frac{\nu\alpha}{1-\nu} \tag{S28}$$

As an example, we consider the room-temperature ionic liquid $EMIM^+TFSI^-$ whose physical properties are given as $\alpha = 11.3$, $\delta\alpha = 0.19$ (Table S2), $D_+ = 5.4 \times 10^{-11} \, m^2 \, s^{-1}$, $D_- = 3.3 \times 10^{-11} \, m^2 \, s^{-1}$ [11] (thus $\chi = 0.24$) and $\lambda = 9.4 \, mS \, cm^{-1}$ [12] (conductivity). With above parameter values, we can estimate the bulk ion strength as $n_0 \approx 1.732 \times 10^{27} \, m^{-3}$ using the Nernst-Einstein relationship $\lambda = e^2 z^2 (D_+ + D_-) n_0 / (k_B T)$. Based on the ion sizes shown in Fig. S2, we can estimate the effective diameters of $EMIM^+$ and $TFSI^-$ as $\sigma_+ = (5 \, Å \times 9.9 \, Å \times 5 \, Å)^{1/3} \approx 5.52 \, Å$ and $\sigma_- = (4.6 \, Å \times 11.5 \, Å \times 4.8 \, Å)^{1/3} \approx 6.33 \, Å$, respectively. Then, we approximate the ion diameter as $\sigma = (\sigma_+ + \sigma_-)/2 \approx 6 \, Å$ and obtain $\nu = 2\sigma^3 n_0 \approx 0.75$. Substituting these parameter values into (S28), we obtain $S_e^{est} \approx -0.69 \, mV \, K^{-1}$, which agrees with the experimental result [13] in both sign and order of magnitude. However, the predicted Seebeck coefficient is relatively small compared to the experimental value ($S_e = -0.85 \, mV \, K^{-1}$ [13]). This could be due to the inaccurate estimation of $\nu$ caused by the underestimation of $\sigma$ (Kilic et al. [14] argued that $\sigma$ is "*a cutoff for the unphysical divergences of PB theory*" and is increased by solvent effects and ion correlations). Conversely, from the experimental data $S_e = -0.85 \, mV \, K^{-1}$, we can infer the effective ion volume fraction as $\nu \approx 0.79$, which is only about 5% higher than the estimated value ($\sim 0.75$). Therefore, in Fig. 3 we take $\nu$ as $\nu = 0.79$ and find that $S_e \to -9.95 k_B/e \approx -0.85 \, mV \, K^{-1}$ as $\kappa_{nom} H \to \infty$ (i.e., nominal Debye length $\kappa_{nom}^{-1} \ll H$).

It is worth mentioning that our model (Eq. 5) does not include the thermally induced electrode potential difference that is not eliminated from other experiments [15, 16]. Therefore, a direct comparison between our theoretical prediction and these experiments is not feasible unless the electrode potential difference is incorporated into the theoretical model or is subtracted from the experimental data. This will be considered in our future investigation.

## S4. DETAILS OF NUMERICAL SIMULATION

For convenience, the ionic flux can be rewritten as

$$\boldsymbol{J}_i^{\mathrm{mol}} = -D_i \Gamma_0 \left[ \nabla \tilde{n}_i + \frac{(\nu_0/2)\tilde{n}_i \sum_j \nabla \tilde{n}_j}{1 - (\nu_0/2)\sum_j \tilde{n}_j} + \frac{z_i e \tilde{n}_i}{k_{\mathrm{B}} T} \nabla \phi + \frac{2\tilde{n}_i \alpha_i}{T} \nabla T \right] \tag{S29}$$

Correspondingly, the free charge density can be rewritten as

$$\rho_{\mathrm{e}} = ze\Gamma_0(\tilde{n}_+ - \tilde{n}_-) \tag{S30}$$

Interestingly, with the above transformation, $\tilde{n}_i$ is identical to the corresponding ion molar concentration $c_i$ in magnitude. Then, the governing equations for the investigated system becomes

$$\nabla^2 T = 0 \tag{S31}$$

$$\nabla \cdot (\varepsilon_0 \varepsilon_{\mathrm{r}} \nabla \phi) = -\rho_{\mathrm{e}} \tag{S32}$$

$$\nabla \cdot \boldsymbol{J}_i^{\mathrm{mol}} = 0 \tag{S33}$$

Our numerical simulations were carried out using the commercial finite element software COMSOL Multiphysics. Three built-in interfaces were used in our simulations, i.e., a *Heat Transfer in Fluids* interface for the energy equation [i.e. (S31)], an *Electrostatics* interface for the Poisson equation [i.e. (S32)] and a *General PDE* interface for the modified Nernst-Planck equations [i.e. (S33)]. The computational domain is given by Fig. 4a in the main text and consist of one half of nanochannel (due to symmetry) and two reservoirs connected to its two ends. Such computational domain was the same as that considered in Ref. [6]. Specifically, the nanochannel length was chosen as $200H$ and the reservoir size was set to $20H$ (length) $\times 5H$ (height). For simplicity, the temperature dependence of the relative permittivity ($\varepsilon_{\mathrm{r}} = 78.408$) and ion diffusion coefficients ($D_+ = D_- = 2 \times 10^{-9}\,\mathrm{m^2/s}$) were

neglected in the simulations. In addition, the relative tolerance was set to $10^{-5}$ in all calculations.

The boundary conditions were also displayed in Fig. 4a. The following briefly introduces the boundary conditions used in the simulations.

1) For (S31), to achieve the temperature conditions that the wall temperature of the nanochannel linearly increases from $T_C$ to $T_H$ (here $T_H - T_C = \Delta T$), the temperatures of the leftmost and rightmost boundaries were set to $T_C = T_0 - \Delta T/2$ and $T_H = T_0 + \Delta T/2$, respectively. In addition, for all numerical simulations, we set $T_0 = 298\,\mathrm{K}$ and $\Delta T = 10\,\mathrm{K}$.

TABLE S1. Surface charge density ($\Sigma/\mathrm{mC} \cdot \mathrm{m}^{-2}$) used as a boundary condition to implement numerical simulation of (S33) for varying nominal dimensionless Debye parameters $\kappa_{\mathrm{nom}} H$ with two different volume fraction $\nu_0$ as $\psi_w = -15$ mV and $\psi_w = -75$ mV. The temperature was fixed as 298 K and relative permittivity was set to $\varepsilon_r = 78.408$. In addition, the ion strength was set to $n_0 = \Gamma_0 = 1\,\mathrm{mM} \times N_A$ and thus $\kappa_{\mathrm{nom}}^{-1} \approx 10\,\mathrm{nm}$.

| | $\psi_w = -15$ mV | | $\psi_w = -75$ mV | |
| $\kappa_{\mathrm{nom}} H$ | $\nu_0 = 0$ | $\nu_0 = 0.1$ | $\nu_0 = 0$ | $\nu_0 = 0.1$ |
| --- | --- | --- | --- | --- |
| 0.01 | $-0.011441$ | $-0.011244$ | $-0.17121$ | $-0.093561$ |
| 0.05 | $-0.057151$ | $-0.056167$ | $-0.84976$ | $-0.46686$ |
| 0.1 | $-0.11397$ | $-0.11202$ | $-1.6623$ | $-0.92788$ |
| 0.2 | $-0.22531$ | $-0.22159$ | $-3.0751$ | $-1.8107$ |
| 0.5 | $-0.52254$ | $-0.51547$ | $-5.5349$ | $-3.8959$ |
| 1 | $-0.84677$ | $-0.83969$ | $-6.8818$ | $-5.5369$ |
| 2 | $-1.0593$ | $-1.0542$ | $-7.4494$ | $-6.3096$ |
| 3 | $-1.0918$ | $-1.087$ | $-7.5318$ | $-6.4201$ |
| 5 | $-1.0969$ | $-1.0921$ | $-7.5452$ | $-6.4379$ |
| 10 | $-1.097$ | $-1.0922$ | $-7.5455$ | $-6.4382$ |
| 100 | $-1.097$ | $-1.0922$ | $-7.5455$ | $-6.4382$ |

2) For (S32), we can not impose the constant $\psi_w$ on the solid-solution interfaces directly since the electric potential $\phi$ considered here is the superposition of the EDL potential $\psi$ and the induced potential $\phi_v$. To impose these boundary conditions correctly, it requires to

calculate the corresponding $\Sigma$ as a function of $\psi_w$, $\kappa_{nom}H$ and $\nu_0$, i.e., $\Sigma = \Sigma(\psi_w, \kappa_{nom}H, \nu_0)$. For the cases without ion steric effects, the classic Poisson-Boltzmann equation can be solved analytically to obtain $\Sigma(\psi_w, \kappa_{nom}H)$ [17]. Unfortunately, this approach doses not work for the cases with ion steric effects because the Poisson-Fermi equation can not be analytically solved. Hence, we evaluated the $\Sigma$ based on the numerical solutions of the Poisson-Fermi equation (9). Table S1 lists the calculated $\Sigma$ as a function of $\psi_w$, $\kappa_{nom}H$ and $\nu_0$.

In addition, the potential of the leftmost reservoir boundary was set to zero, while the rightmost reservoir boundary was specified by various values (i.e., parametric study). Subsequently, the induced potential was determined by linear interpolation from the obtained current-voltage characteristics.

3) For (S33), the keypoint is correlately imposing the boundary conditions on the leftmost and rightmost boundaries of the system. As pointed out by Ref. [6], the Soret equilibrium gives

$$\frac{1}{n}\frac{\mathrm{d}n}{\mathrm{d}x} = -\alpha\frac{1}{T}\frac{\mathrm{d}T}{\mathrm{d}x} \tag{S34}$$

Integrating (S34) and taking advantage of $\int_0^L n(x)\mathrm{d}x = n_0 L$, we obtain

$$n(x) = n_0 \left[\frac{T_0}{T(x)}\right]^\alpha \tag{S35}$$

Therefore, we find $n(0) = n_0[T_0/(T_0 - \Delta T/2)]^\alpha$ and $n(L) = n_0[T_0/(T_0 + \Delta T/2)]^\alpha$. These two conditions were imposed on the leftmost and rightmost boundaries of the investigated system.

## S5. COMPARISON OF CONSTANT SURFACE POTENTIAL AND CONSTANT SURFACE CHARGE DENSITY BOUNDARY CONDITIONS

For the cases without ion steric effects, the values of $\Sigma$ used in the studies with a constant surface charge density boundary condition can be directly calculated by the Grahame equation based on the famous Gouy-Chapman model. In the presence of the ion steric effects, the modified Grahame equation [18]

$$\Sigma = \mathrm{sgn}(\psi_w)\sqrt{\frac{4nk_BT\epsilon}{\nu}\ln\left[1 - \nu + \nu\cosh\left(\frac{ze\psi_w}{k_BT}\right)\right]} \tag{S36}$$

$$= \mathrm{sgn}(\psi_w)\frac{ezn}{\kappa}\sqrt{\frac{8}{\nu}\ln\left[1 - \nu + \nu\cosh\left(\frac{ze\psi_w}{k_BT}\right)\right]} \tag{S37}$$

is used to relate the surface charge density and surface potential. Clearly, as $\nu \to 0$, (S36) tends to the famous Grahame equation

$$\Sigma = \sqrt{8nk_{\mathrm{B}}T\epsilon} \sinh\left(\frac{ze\psi_{\mathrm{w}}}{2k_{\mathrm{B}}T}\right) = \frac{4ezn}{\kappa}\sinh\left(\frac{ze\psi_{\mathrm{w}}}{2k_{\mathrm{B}}T}\right) \tag{S38}$$

In the calculations, $n_0 = \Gamma_0 = 1\,\mathrm{mM} \times N_{\mathrm{A}}$ and thus $\kappa_{\mathrm{nom}}^{-1} \approx 10\,\mathrm{nm}$. Equation (S37) is valid for nonoverlapping EDLs and was used to estimate the the values of $\Sigma$ for the constant surface charge boundary condition corresponding to $\psi_{\mathrm{w}}$. Note that the constant $\Sigma$ may lead to unphysically large prediction for $S_{\mathrm{e}}$ as $\kappa_{\mathrm{nom}}H \to 0$ and give rise to the same prediction as the constant surface potential bounary condition as $\kappa_{\mathrm{nom}}H \to \infty$ (see details in the main text).

## S6. RATIONALITY OF REDUCED SORET COEFFICIENTS USED IN THE CALCULATIONS

Table S2 shows the experimental values for the ion activation enthalpies $\Delta H_{\pm}$ and corresponding derived values of the ion reduced Soret coefficents $\alpha_{\pm}$ for both simple salts [9] (such as NaCl and KCl) and ionic liquids [11] (such as EMIM$^+$TFSI$^-$ and EMIM$^+$OAc$^-$). For simple salts, we use the values of the ion reduced Soret coefficients $\alpha_{\pm}$ from Ref. [9], which were evaluated from experimental values of ion heats of transport [4]. For ionic liquids, we calculate the values of the ion reduced Soret coefficients $\alpha_{\pm}$ by $\alpha_{\pm} = Q_{\pm}/(2k_{\mathrm{B}}T)$, where $Q_{\pm}$ given by [19]

$$Q_{\pm} = k_{\mathrm{B}}T + \Delta H_{\pm} \tag{S39}$$

with $\Delta H_{\pm}$ being the ion activation enthalpy.

As seen from Table S2, for the ionic liquid EMIM$^+$OAc$^-$, $\alpha_{\pm}$ can reach up to $\sim 8$, so it is reasonable to consider smaller $\alpha_{\pm}$ values (e.g., $\alpha_{\pm} = 5$) in this study. Moreover, for the ionic liquid EMIM$^+$TFSI$^-$, we find that $\alpha_+ \approx \alpha_-$ is very close to $\sim 5$. On the other hand, we assume $\alpha_+ = \alpha_-$ to focus on the new thermoelectric mechanism and neglect the classic thermoelectric effect due to the difference in reduced Soret coefficients between cations and anions.

TABLE S2. Activation enthalpy, reduced Soret coefficients and other derived values of dissolved ions in various electrolytes.

| | Simple salts | | Ionic liquids | |
|---|---|---|---|---|
| Electrolyte | NaCl | KCl | EMIM$^+$TFSI$^{-}$[a] | EMIM$^+$OAc$^{-}$[b] |
| $\Delta H_+$/eV | / | / | 0.27 [11] | 0.36 [11] |
| $\Delta H_-$/eV | / | / | 0.26 [11] | 0.39 [11] |
| $\alpha_+$ | 0.7 [9] | 0.5 [9] | 5.76[c] | 7.51[c] |
| $\alpha_-$ | 0.1 [9] | 0.1 [9] | 5.56[c] | 8.09[c] |
| $\alpha = \alpha_+ + \alpha_-$ | 0.8 | 0.6 | 11.3 | 15.6 |
| $\delta\alpha = \alpha_+ - \alpha_-$ | 0.6 | 0.4 | 0.19 | −0.58 |

[a] 1-ethyl-3-methylimidazolium-bis(trifluoro-methylsulfonyl)imide.

[b] 1-ethyl-3-methylimidazolium acetate.

[c] Calculated by $\alpha_i = 0.5 + \Delta H_i/(2k_{\mathrm{B}}T)$ [19].

## S7.  PERSPECTIVE IN THERMOELECTRICICITY OF CONFINED ELECTROLYTES USING MOLECULAR DYNAMICS SIMULATION

Over the past decade, the (nonequilibrium) molecular dynamics simulation (MD) was successfully applied to investigate the thermoosmosis in nanofluidic systems [20, 21]. This approach can be further extended to study the thermoelectric responses in nanofluidic systems. In what follows, we briefly introduce how to implement it.

The nonequilibrium thermodynamics theory suggests that the electric current density $j_{\mathrm{e}}$ and and the heat flux density $j_{\mathrm{h}}$ are linked with the external gradients of electric potential $-\nabla\phi$ and temperature $-\nabla T$ by [22, 23]

$$\begin{bmatrix} j_{\mathrm{e}} \\ j_{\mathrm{h}} \end{bmatrix} = \begin{bmatrix} \lambda & M_{12} \\ M_{21} & kT \end{bmatrix} \begin{bmatrix} -\nabla\phi \\ -\nabla T/T \end{bmatrix} \tag{S40}$$

where where $\lambda$ and $k$ are the electrical and the thermal conductivities of the nanofluidic system; $M_{ij}$ are the phenomenological coefficients and the Onsager relation suggests $M_{12} = M_{21} = M_{\mathrm{TE}}$.

The Seebeck coefficient can be evaluated by $S_e = M_{\mathrm{TE}}/(kT)$ [22, 23]. Therefore, there are two routes that can be used to calculate the Seebeck coefficient: one is the electrocaloric

route (i.e., calculating $M_{21}$) and the other is the thermoelectric route (i.e., calculating $M_{12}$). For the former, one considers electroosmotic flow in nanofluidic systmes and computes the heat flux as [20–23]

$$j_{\mathrm{h}} = \frac{1}{2H} \int_{-H}^{H} \delta h(y) v(y) \mathrm{d}y \tag{S41}$$

with $\delta h$ the excess enthalpy density and $v$ the velocity. Then, the coefficient $M_{21}$ is computed by $M_{21} = j_{\mathrm{h}}/(-\nabla\phi)$. The calculations of $\delta h$ and $v$ have been detailed in the literature [22, 23]. For the latter, the investigated system should be the same as that in Ref. [20, 21]. The key of this route is computing the current density $j_{\mathrm{e}}$ from the time evolution of the charge carriers in the reservoirs. Subsequently, one can compute $M_{12} = j_{\mathrm{e}}/(-\nabla T/T)$. This is beyond the scope of this study and may be left for future investigations.

---

[1] H. B. Callen, The application of onsager's reciprocal relations to thermoelectric, thermomagnetic, and galvanomagnetic effects, Phys. Rev. **73**, 1349 (1948).

[2] X. Qian, T.-H. Liu, and R. Yang, Confinement effect on thermopower of electrolytes, Mater. Today Phys. **23**, 100627 (2022).

[3] M. S. Kilic, M. Z. Bazant, and A. Ajdari, Steric effects in the dynamics of electrolytes at large applied voltages. ii. modified poisson-nernst-planck equations, Phys. Rev. E **75**, 021503 (2007).

[4] J. N. Agar, C. Y. Mou, and J. L. Lin, Single-ion heat of transport in electrolyte solutions: a hydrodynamic theory, J. Phys. Chem. **93**, 2079 (1989).

[5] I. Borukhov, D. Andelman, and H. Orland, Steric effects in electrolytes: a modified poisson-boltzmann equation, Phys. Rev. Lett. **79**, 435 (1997).

[6] M. Dietzel and S. Hardt, Thermoelectricity in confined liquid electrolytes, Phys. Rev. Lett. **116**, 225901 (2016).

[7] P. B. Peters, R. van Roij, M. Z. Bazant, and P. M. Biesheuvel, Analysis of electrolyte transport through charged nanopores, Phys. Rev. E **93**, 053108 (2016).

[8] S. Alizadeh and A. Mani, Multiscale model for electrokinetic transport in networks of pores, part i: Model derivation, Langmuir **33**, 6205 (2017).

[9] A. Würger, Thermal non-equilibrium transport in colloids, Rep. Prog. Phys. **73**, 126601 (2010).

[10] F. Baldessari, Electrokinetics in nanochannels: Part i. electric double layer overlap and channel-to-well equilibrium, J. Colloid Interface Sci. **325**, 526 (2008).

[11] C. D'Agostino, M. D. Mantle, C. L. Mullan, C. Hardacre, and L. F. Gladden, Ddiffusion, ion pairing and aggregation in 1-ethyl-3-methylimidazolium-based ionic liquids studied by [1]H and [19]F PFG NMR: Effect of temperature, anion and glucose dissolution, ChemPhysChem **19**, 1081 (2018).

[12] J. Cho, J. Lee, Y. He, B. Kim, T. Lodge, and C. Frisbie, High-capacitance ion gel gate dielectrics with faster polarization response times for organic thin film transistors, Adv. Mater. **20**, 686 (2008).

[13] D. Zhao, A. Martinelli, A. Willfahrt, T. Fischer, D. Bernin, Z. U. Khan, M. Shahi, J. Brill, M. P. Jonsson, S. Fabiano, and X. Crispin, Polymer gels with tunable ionic seebeck coefficient for ultra-sensitive printed thermopiles, Nat. Commun. **10** (2019).

[14] M. S. Kilic, M. Z. Bazant, and A. Ajdari, Steric effects in the dynamics of electrolytes at large applied voltages. i. double-layer charging, Phys. Rev. E **75**, 021502 (2007).

[15] T. Li, X. Zhang, S. D. Lacey, R. Mi, X. Zhao, F. Jiang, J. Song, Z. Liu, G. Chen, J. Dai, Y. Yao, S. Das, R. Yang, R. M. Briber, and L. Hu, Cellulose ionic conductors with high differential thermal voltage for low-grade heat harvesting, Nat. Mater. **18**, 608 (2019).

[16] C.-G. Han, X. Qian, Q. Li, B. Deng, Y. Zhu, Z. Han, W. Zhang, W. Wang, S.-P. Feng, G. Chen, and W. Liu, Giant thermopower of ionic gelatin near room temperature, Science **368**, 1091 (2020).

[17] S. H. Behrens and M. Borkovec, Exact poisson-boltzmann solution for the interaction of dissimilar charge-regulating surfaces, Phys. Rev. E **60**, 7040 (1999).

[18] M. Z. Bazant, M. S. Kilic, B. D. Storey, and A. Ajdari, Towards an understanding of induced-charge electrokinetics at large applied voltages in concentrated solutions, Adv. Colloid Interface Sci. **152**, 48 (2009).

[19] A. Würger, Thermoelectric ratchet effect for charge carriers with hopping dynamics, Phys. Rev. Lett. **126**, 068001 (2021).

[20] L. Fu, S. Merabia, and L. Joly, What controls thermo-osmosis? molecular simulations show the critical role of interfacial hydrodynamics, Phys. Rev. Lett. **119**, 214501 (2017).

[21] L. Fu, S. Merabia, and L. Joly, Understanding fast and robust thermo-osmotic flows through carbon nanotube membranes: Thermodynamics meets hydrodynamics, J. Phys. Chem. Lett.

**9**, 2086 (2018).

[22] L. Fu, L. Joly, and S. Merabia, Giant thermoelectric response of nanofluidic systems driven by water excess enthalpy, Phys. Rev. Lett. **123**, 138001 (2019).

[23] Y. Jin, R. Tao, S. Luo, and Z. Li, Size-sensitive thermoelectric properties of electrolyte-based nanofluidic systems, J. Phys. Chem. Lett. **12**, 1144 (2021).

[24] M. A. Gebbie, H. A. Dobbs, M. Valtiner, and J. N. Israelachvili, Long-range electrostatic screening in ionic liquids, Proc. Natl. Acad. Sci. **112**, 7432 (2015).

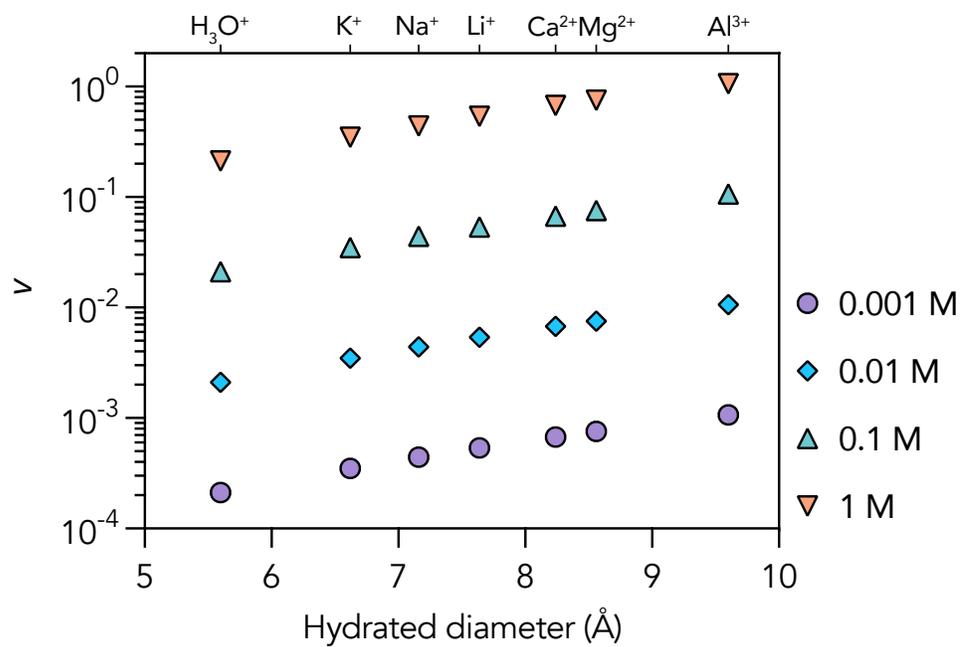

FIG. S1. Variation of ion volume fraction with hydrated diameter for various concentrations.

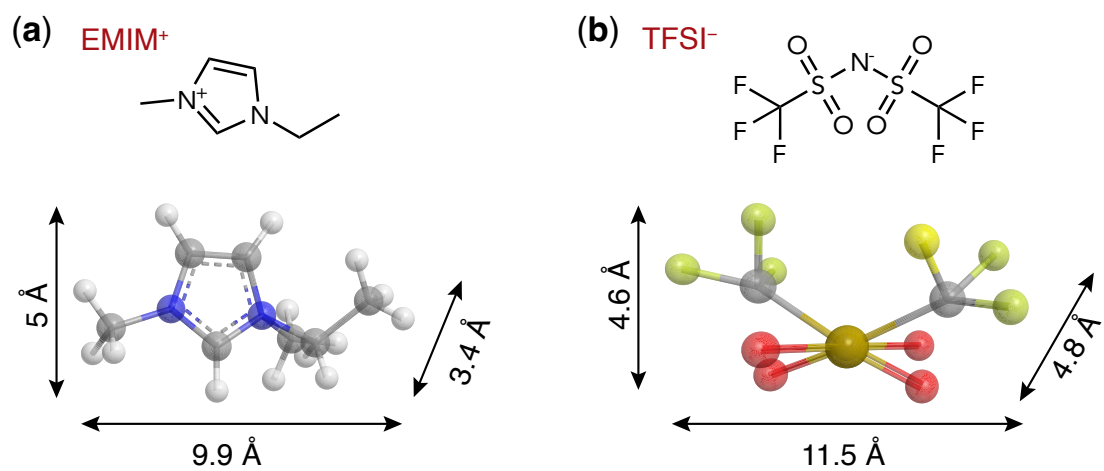

FIG. S2. Chemical structure and approximated dimensions of ionic liquid ions. (a) EMIM$^+$ and (b) TFSI$^-$. Molecular dimensions representing the van der Waals radii of the constituent atoms are taken from Ref. [24].

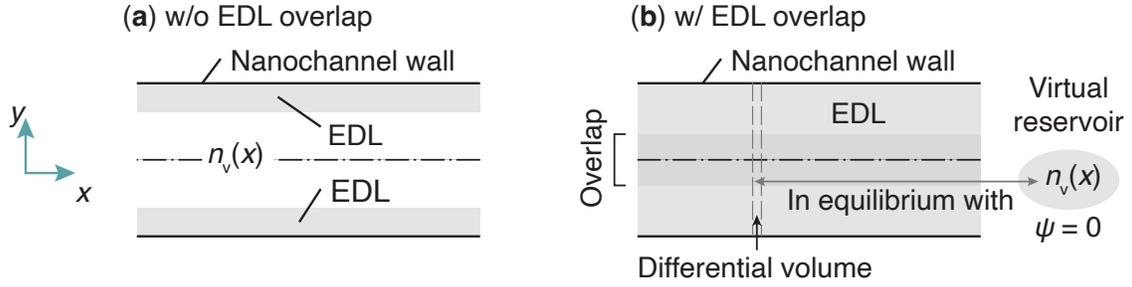

FIG. S3. Illustration of virtual concentration of either ion species $n_{\mathrm{v}}$. (a) Nanochannel without EDL overlap. In this case, $n_{\mathrm{v}}$ represents the ionic strength or number concentration of either ion species at the center plane of the nanechannel. (b) Nanochannel with EDL overlap. In this case, $n_{\mathrm{v}}$ represents the ionic strength or number concentration of either ion species in the virtual electroneutral reservoir ($\psi = 0$) that is in equilibrium with any differential volume in the channel or local cross section [7, 8].



\end{document}